\documentclass[reprint, amsmath,amssymb,aps, onecolumn]{revtex4-2}

\usepackage{graphicx}
\usepackage{subfigure}
\usepackage{subfloat}
\usepackage{amssymb}
\usepackage{amsmath}
\usepackage{multirow}
\usepackage{dcolumn}
\usepackage{bm}
\usepackage{wasysym}
\usepackage{cancel}

\usepackage{txfonts}
\usepackage{lineno}

\begin{document}

\title{$\alpha$ Effect and Magnetic Diffusivity $\beta$ in Helical Plasma under Turbulence Growth}

\author{Kiwan Park}
 \affiliation{ Institute of Plasmas Turbulence and Magnetic Fields(ITPM); Soongsil University, 369, Sangdo-ro, Dongjak-gu, Seoul 06978 Republic of Korea\\ pkiwan@ssu.ac.kr; pkiwan93@naver.com\\}

\date{\today}

\begin{abstract}
We investigate the transport coefficients $\alpha$ and $\beta$ in plasma systems with varying Reynolds numbers while maintaining a unit magnetic Prandtl number. {The $\alpha$ and $\beta$ tensors parameterize the turbulent electromotive force (EMF) in terms of the large-scale magnetic field ${\bf \overline{B}}$ and current density  ${\bf \overline{}}$ as follows : $\langle {\bf u}\times {\bf b} \rangle = \alpha {\bf \overline{B}}-\beta {\nabla\times \bf \overline{B}}$.}  In astrophysical plasmas, high fluid Reynolds numbers ($Re$) and magnetic Reynolds numbers ($Re_\mathrm{M}$) drive turbulence, where $Re$ governs flow dynamics and $Re_\mathrm{M}$ controls magnetic field evolution. {The coefficients $\alpha_{\text{semi}}$ and $\beta_{\text{semi}}$ are obtained from large-scale magnetic field data as estimates of the $\alpha$ and $\beta$ tensors, while $\beta_{\text{theo}}$ is derived from turbulent kinetic energy data.} The reconstructed large-scale field $\overline{B}$ agrees with simulations, confirming consistency among $\alpha$, $\beta$, and $\overline{B}$ in weakly nonlinear regimes. This highlights the need to incorporate magnetic effects under strong nonlinearity. To clarify $\alpha$ and $\beta$, we introduce a field structure model, identifying $\alpha$ as the electrodynamic induction effect and $\beta$ as the fluid-like diffusion effect. The agreement between our method and direct simulations suggests that plasma turbulence and magnetic interactions can be analyzed using fundamental physical quantities. Moreover, $\alpha_{\text{semi}}$ and $\beta_{\text{semi}}$, which successfully reproduce the numerically obtained magnetic field, provide a benchmark for future theoretical studies.
\end{abstract}

\maketitle

\section{Introduction}
Celestial systems, predominantly in a plasma state or surrounded by plasma, are significantly influenced by magnetic fields ($B$). Since plasma consists of charged particles such as electrons and ions, it is highly responsive to magnetic fields. For instance, turbulent plasma energy is converted into magnetic energy through a process known as the dynamo. Conversely, magnetic energy is converted into the plasma's kinetic or thermal energy, as observed in phenomena such as Alfvén waves and magnetic reconnection. {An Alfv\'{e}n wave can induce magnetic reconnection by destabilizing the current sheet. Conversely, the reconnection process can generate Alfv\'{e}n waves, which transport energy throughout the plasma system---a mechanism believed to play a role in solar flares. This coupling process is known to occur when the width of the current sheet becomes comparable to the ion skin depth,
\[
d_i \sim \frac{c}{\omega_{pi}} = c \sqrt{ \frac{\varepsilon_0 m_i}{n_i e^2} }.
\]}
Additionally, the balance between plasma pressure and magnetic pressure determines the stability of the plasma system. Moreover, by transporting angular momentum, magnetic fields act as a brake on celestial plasma systems, gradually slowing their rotation and playing a critical role in the regulation of star formation and the development of accretion disks. Furthermore, magnetorotational instability (MRI) in accretion disks enhances turbulence, driving angular momentum outward and allowing material to accrete inward. These mechanisms, which are crucial to the dynamics of astrophysical systems, are essential for understanding their behavior not only in distant celestial objects but also in the Sun \citep{1991ApJ...376..214B, 1998pfp..book.....C, 2003Phpl.book.....B, 2003dysu.book..217P, 2005MNRAS.362..369M, 2014ARA&A..52..251C}.\\

In addition to these macroscale effects, magnetic fields impact the production of elementary particles. {Perturbed electrons, influenced by the Lorentz force, can be shared among neighboring nuclei, leading to an increase in density due to superposition effects.} Consequently, the enhanced shielding effect reduces the potential barrier between interacting nuclei, thereby elevating nucleosynthesis reaction rates \citep{2024PhRvD.109j3002P}. Since ubiquitous magnetic fields have been present since shortly after the Big Bang and can influence plasma systems while maintaining electrical neutrality, they are among the most probable candidates for contributing to matter production in the Universe.\\

Despite these broad influences of ever-present magnetic fields, their origin is not clearly understood. In the very early Universe, various quantum fluctuation processes, such as QCD or phase transitions, are thought to have formed primordial magnetic fields (PMF) \citep{1950ZNatA...5...65B, 1994PhRvD..50.2421C, 2012ApJ...759...54T}. As the density of plasma particles increased through nucleosynthesis, {Their enhanced collision frequency and plasma fluctuations led to misalignments between the gradients of electron density, $\nabla n_e$, and pressure, $\nabla p_e$ (or temperature, $\nabla T_e$), such that $\nabla p \times \nabla n \neq 0$. Known as the Biermann battery effect, this phenomenon acts as a source term in the magnetic induction equation, generating magnetic fields.} \citep{1950ZNatA...5...65B, 2014NatPh..10..520M}. The primordial magnetic fields (PMF) from this epoch are estimated to have been extremely weak, in the range of $10^{-62}\,\text{G}-10^{-19}\,\text{G}$, compared to the currently observed average magnetic field strength $10^{-9}\,\text{G}-10^{-5}\,\text{G}$ \citep{2012PhR...517..141Y}. This discrepancy suggests that more enhanced and efficient dynamo processes must have existed \citep{2005PhR...417....1B, 2012MNRAS.419..913P, 2012MNRAS.423.2120P}.\\

The generation, amplification, and propagation of magnetic fields in plasmas differ significantly from those in free space. In free space, the induction and propagation of magnetic fields can be explained by Maxwell's equations and electrodynamics. {However, in plasmas, the presence of massive charged particles and fluctuating magnetic structures mutually interfere, giving rise to additional magnetic field structures.} {Often, the induced magnetic eddy remains a local fluctuation that decays into smaller-scale structures rather than evolving or propagating to larger scales, due to the interactive constraints imposed by other massive charged particles. Therefore, the generation and sustained propagation of magnetic fields in plasmas require additional energy and constraints. The behavior of magnetic fragments in plasmas resembles that of fluid eddies.}\\

The transport of magnetic fields in plasmas, leading to their amplification at large or small scales, is referred to as dynamo \citep{2005PhR...417....1B}. These phenomena are essentially the induction of the magnetic field through electromotive force (EMF, $\mathbf{U} \times \mathbf{B}$, where $\mathbf{U}$ represents fluid velocity). The migration of magnetic energy toward larger scales is referred to as an inverse cascade of energy, leading to a large-scale dynamo (LSD). {Conversely, the transfer of magnetic energy toward smaller scales, resulting in a peak in the small scale regime, is referred to as a small-scale dynamo (SSD).} The forward cascade of energy is often observed in both hydrodynamics (HD) and magnetohydrodynamics (MHD). However, the inverse cascade requires more stringent conditions, such as 2D HD systems with conserved kinetic energy $\langle U^2 \rangle $ and enstrophy {$\langle (\nabla \times \mathbf{U})^2\rangle$}, 3D MHD systems with helicity (\(\langle \mathbf{A}\cdot \mathbf{B}\rangle \neq 0,\,\, \nabla \times \mathbf{A} = \mathbf{B}\)), differential rotation, or magnetorotational instability (MRI), also known as the Balbus-Hawley instability \citep{1966ZNatA..21.1285S, 2001ApJ...550..824B, 1991ApJ...376..214B}. The migration and amplification of the magnetic field are influenced by various factors, and the critical conditions governing these processes remain an active area of research and debate. In particular, the turbulent electromotive force, which acts as a source for the large-scale magnetic field, is {not easy to analyze its physical theoretical properties.}\\ With helicity in magnetic fields, the turbulent electromotive force \(\langle \mathbf{u} \times \mathbf{b} \rangle\) can be represented as \(\alpha\overline{\mathbf{B}} - \beta \nabla \times \overline{\mathbf{B}}\), regardless of whether the system is dominated by kinetic or magnetic energy. {Here, $\alpha$ and $\beta$ are pseudo and regular tensors, respectively, and $\overline{\bf B}$ denotes the large-scale magnetic field.}\\

The \(\alpha\) and \(\beta\) coefficients not only simplify the nonlinear dynamo process but also allow an algebraic approach to the dynamo process for further theoretical investigation. {In particular, for rotational structures such as the Sun, \(\alpha\) and \(\beta\) with the angular velocity $\Omega$ describe the periodically oscillating magnetic fields} in the northern and southern hemispheres and the evolution of the poloidal and toroidal fields with a phase difference of 90 degrees \citep{1982GAM....21.....P, 2014ARA&A..52..251C}. {Moreover, the solar magnetic induction equation, incorporating the full forms of $\alpha$ and $\beta$ along with the large-scale poloidal and toroidal magnetic fields $\overline{B}_\mathrm{pol}$, $\overline{B}_\mathrm{tor}$, may explain not only the 22-year periodic evolution of the solar magnetic field but also non-periodic solar (magnetic) activities.} There have been numerous efforts to determine these coefficients through dynamo theories such as mean field theory (MFT), the eddy-damped quasi-normal Markovian (EDQNM) approximation, or the direct interaction approximation (DIA) \citep{1966ZNatA..21.1285S, 1976JFM....77..321P, Akira2011}. However, only approximate forms of the \(\alpha\) and \(\beta\) coefficients are available at present. These theories suggest that \(\alpha\) is composed of residual helicity, \(\langle \mathbf{b} \cdot (\nabla \times \mathbf{b}) \rangle - \langle \mathbf{u} \cdot (\nabla \times \mathbf{u}) \rangle\), while \(\beta\) is related to turbulent energy \(\langle u^2 \rangle\) or \(\langle b^2 \rangle\). It has long been considered that \(\alpha\) induces magnetic fields, while \(\beta\), combined with molecular resistivity \(\eta\), was thought to simply diffuse the fields. {However, their derivations represent first- and second-order approximations that involve unresolved integral factors or Green functions. Since these cannot be solved self-consistently, there is reason to question whether such tentative conclusions, based on incomplete analytical approaches, are indeed valid.}\\

From the viewpoint of magnetohydrodynamics (MHD), the net diffusivity consists of the molecular diffusivity \(\eta\) and the turbulence diffusivity \(\beta\). Similar to \(\alpha\) and \(\beta\), \(\eta\) is also a tensor that is reciprocally related to the conductivity \(\sigma\). The characteristics of \(\eta\) depend on both temperature and density. More critically, \(\eta\) is characterized by the frequency \(\omega\) and the magnetic field \(B\). \(\eta\) can take on negative values depending on \(\omega\) and \(\omega_\mathrm{cs}\). {When $\eta$ is analyzed using kinetic models, this tensor has the potential to become negative depending on the wave frequency $\omega$ and {the cyclotron frequency $\omega_\mathrm{cs} (= qB/2\pi m$, q: charge, m: mass of the particle)}. Interestingly, MHD fluid simulations of decaying magnetized plasma systems also demonstrate the inverse transport of magnetic energy due to negative diffusion, regardless of the conditions typically associated with inverse cascade} \citep{2013PhRvE..87c3002M, 2015PhRvL.114g5001B, 2017MNRAS.472.1628P}. However, the inverse transport of magnetic energy in the decaying system is more related to turbulent diffusivity rather than molecular properties. In this article, we investigate the negative turbulence diffusivity, the \(\beta\) effect in the helically forced system. This highlights the fluid dominance over the electromagnetic \(\alpha\) effect in plasmas composed of many charged particles.\\

To determine the magnetic diffusivity, the fluctuation in \(\alpha\) was considered. Kraichnan proposed that if the fluctuation of \(\alpha\) were large enough, the diffusion could become negative \cite{1976JFM....75..657K}. This model was further generalized by Moffatt, followed by numerous analytical and numerical models \citep{1997ApJ...475..263V, 2014MNRAS.445.3770S}. Later, a more direct numerical method to calculate the \(\alpha\) and \(\beta\) effects was proposed. Using an additional test magnetic field \(\mathbf{B}_T\), fluctuating magnetic fields \(\mathbf{b}_T\) were obtained numerically and used for the computation of dynamo coefficients \citep{2005AN....326..245S, 2018JPlPh..84d7304B}. Recently, Bendre et al.\cite{2024MNRAS.530.3964B} proposed the iterative removal of sources (IROS) method, which uses the time series of the mean magnetic field and current as inputs. Besides, negative magnetic diffusivity was also observed in a liquid sodium experiment \citep{2014PhRvL.113r4501C}, where small-scale turbulent fluctuations contributed to negative magnetic diffusivity in the interior region.\\

In our previous work \citep{2023ApJ...944....2P, 2025PhRvD.111b3021P}, we derived \(\alpha_{\text{semi}}\) and \(\beta_{\text{semi}}\) using large-scale magnetic energy \(\overline{E}_\mathrm{M}(= \langle \overline{B}^2 \rangle /2)\) and magnetic helicity \(\overline{H}_\mathrm{M}(= \langle \overline{A} \cdot \overline{B} \rangle, ~ \overline{B} = \nabla \times \overline{A})\) without {any additional assumptions} or the need to impose extra constraints on the system. We generated their profiles using raw simulation data and successfully reproduced the evolving large-scale magnetic field, consistent with direct numerical simulations (DNS). After confirming that the model produced results in agreement with DNS, we derived a new form of \(\beta_{\text{theo}}\) that requires turbulent kinetic energy \(\langle u^2 \rangle /2\) and helicity \(\langle u \cdot \nabla \times u \rangle\). We constructed the profile of \(\beta_{\text{theo}}\) and reproduced the large-scale magnetic field, again in good agreement with other results. \(\alpha_{\text{semi}}\) and \(\beta_{\text{semi}}\) thus serve as effective references, providing a standard for evaluating new methodologies.\\

In this article, we applied the model to systems with increased magnetic Reynolds numbers (\(Re_\mathrm{M}\)), indicating enhanced nonlinearity but keeping a magnetic Prandtl number (\(Pr_\mathrm{M} = 1\)). We found that the model produced consistent results, particularly in the kinematic regime. Then, we applied the turbulent kinetic data to the new form of \(\beta\). The extended {\(\beta_{\mathrm{theo}}\)} model also successfully reproduced the large-scale magnetic field. This result indicates that turbulent kinetic helicity—previously neglected in conventional theory—plays a crucial role in magnetic diffusivity. Since our model does not rely on artificial assumptions or well-contrived methods, it can be applied not only to DNS but also to observational data. This is what sets our model apart from others. However, the application of our model to the increasing $Re_\mathrm{M}$ system also demonstrates that there is a room for improvement. The increasing magnetic effect with \(Re_\mathrm{M}\) needs to be included. {Additionally, the results indicated the need for an in-depth study of the physical meaning of the correlation length `\(l\)'.}\\

The structure of this paper is as follows. In Section~2, we describe the numerical model and code. Section~3 presents the results of numerical calculations, while Section~4 explains the related theory. For consistency and readability, the theory is not simply cited but is presented with additional detailed explanations. We add a field structure model to elucidate the intuitive and physical meaning of $\alpha$ and $\beta$. We also include a partial IDL script used for calculating the large scale magnetic field $\overline{B}$. Using a field structure model, we explain the physical meanings of \(\alpha\), \(\beta\), and magnetic diffusion $\nabla^2\overline{B}$ in the induction of the magnetic field, considering both nonhelical and helical velocity field structures. {The final section} provides a summary.

\begin{figure*}
    {
   \subfigure[]{
     \includegraphics[width=9.2 cm]{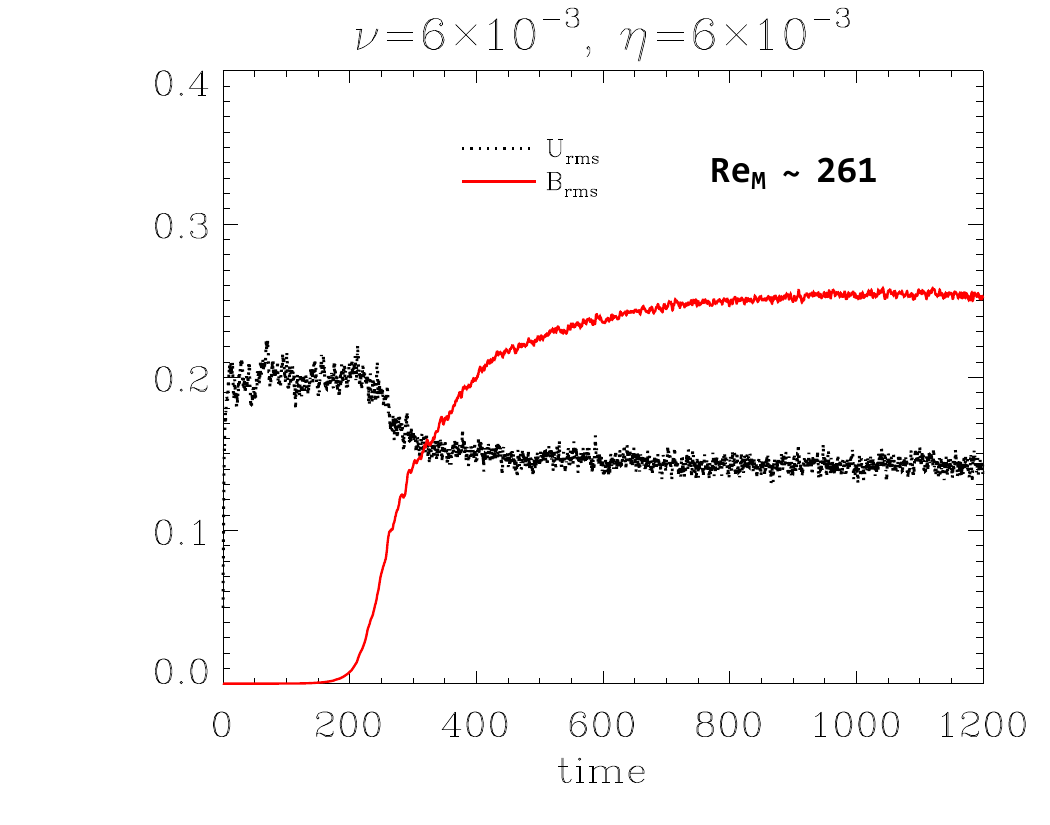}
     \label{f1a}
    }\hspace{-13 mm}
   \subfigure[]{
   \includegraphics[width=9.2 cm]{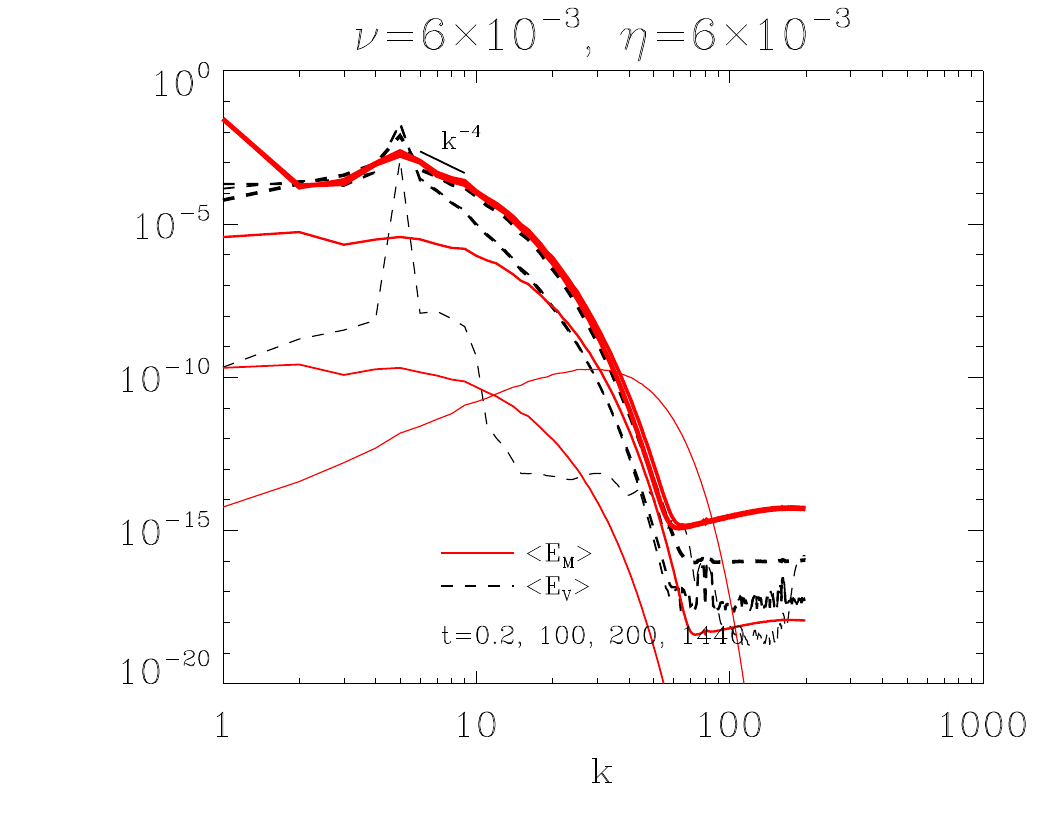}
     \label{f1b}
   }\hspace{-13 mm}
   \subfigure[]{
   \includegraphics[width=9.2 cm]{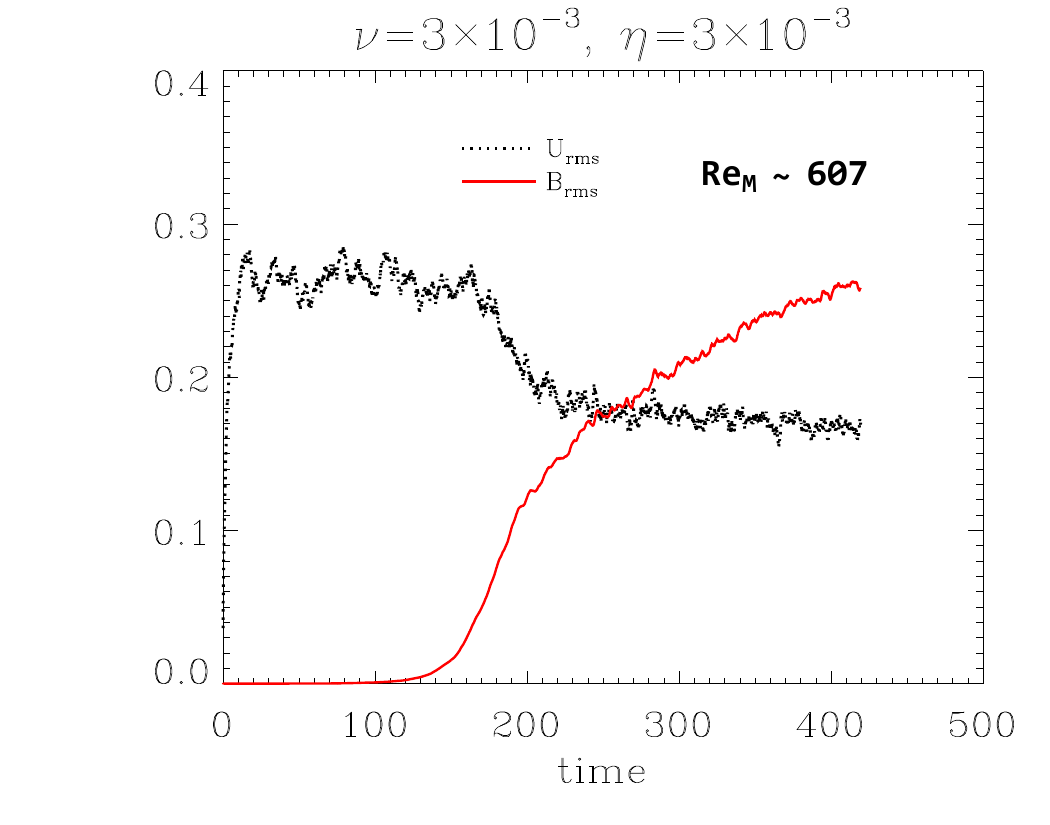}
     \label{f1c}
   }\hspace{-13 mm}
   \subfigure[]{
     \includegraphics[width=9.2 cm]{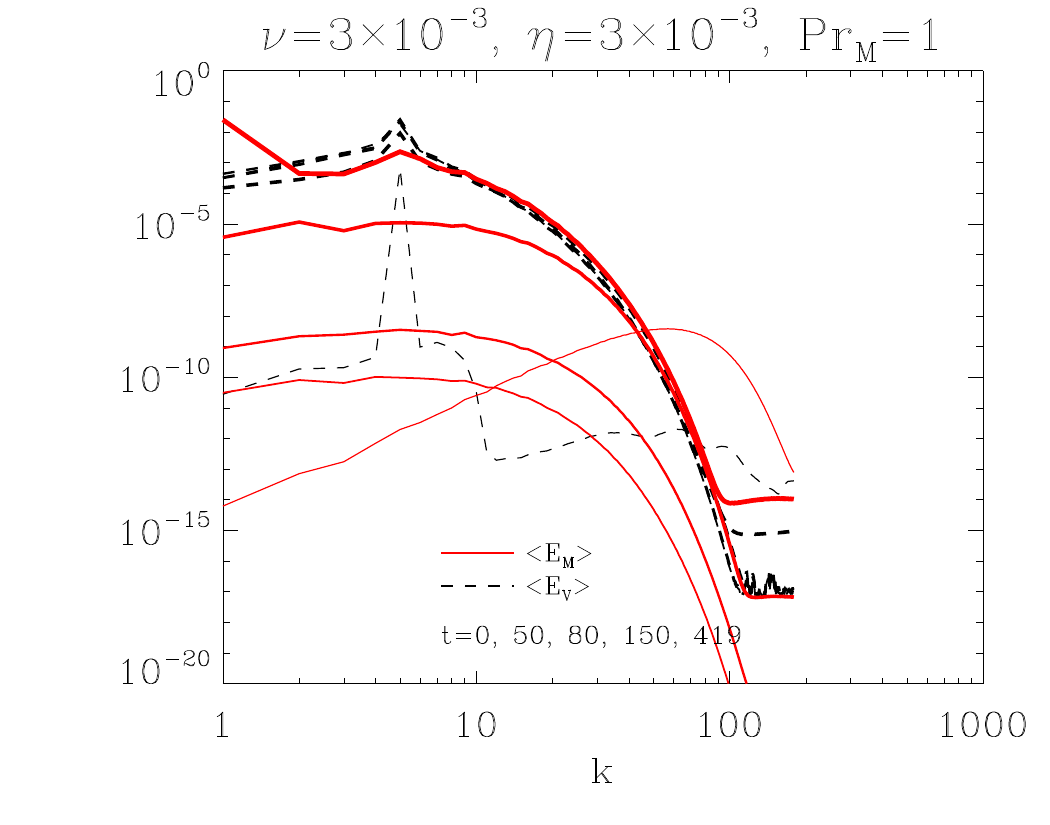}
     \label{f1d}
   }\hspace{-13 mm}
      \subfigure[]{
   \includegraphics[width=9.2 cm]{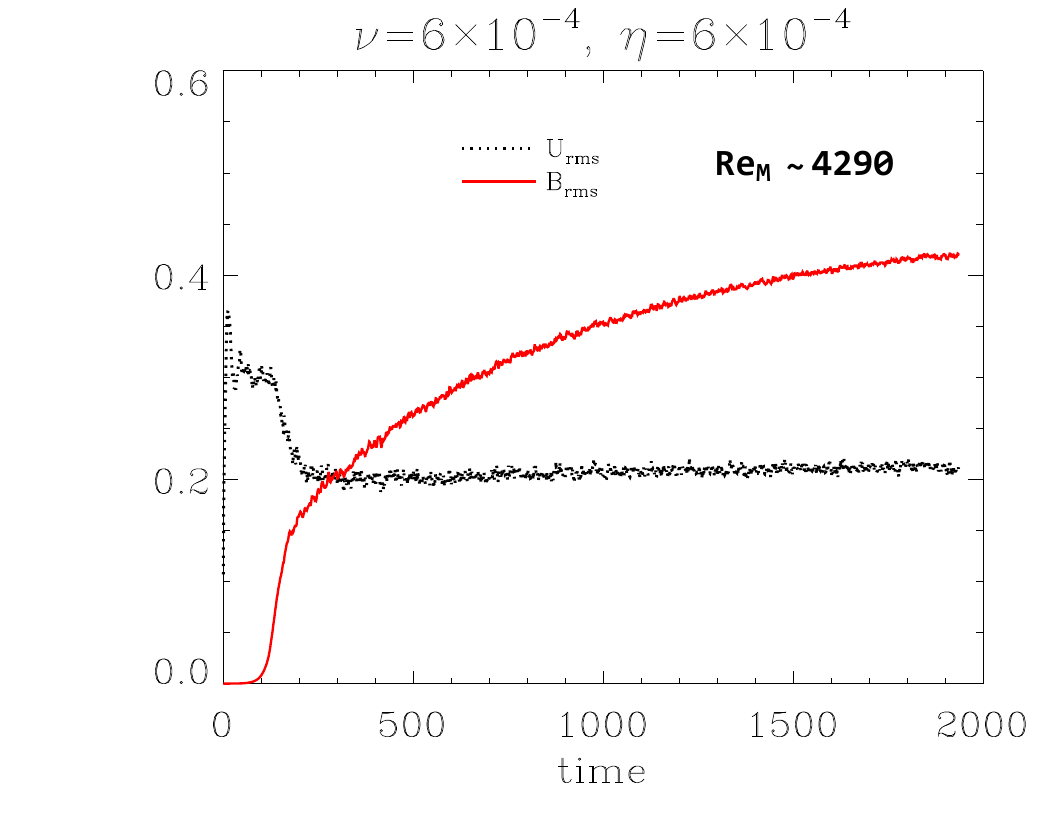}
     \label{f1e}
   }\hspace{-13 mm}
   \subfigure[]{
     \includegraphics[width=9.2 cm]{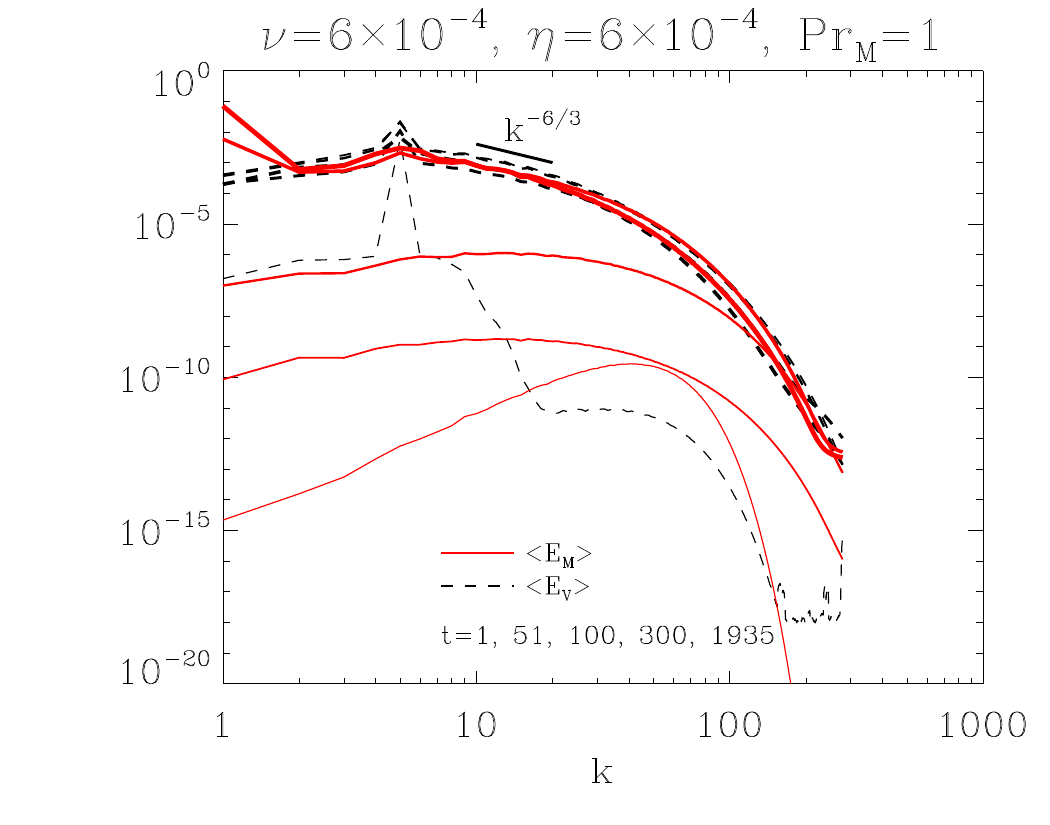}
     \label{f1f}
   }
}
\caption{\raggedright The slopes $k^{-4}$ and $k^{-6/3}$ in (b) and (f) indicate the energy scaling behavior. Note that as $Re_M$ increases, the slope approaches the Kolmogorov's result $-5/3$.}
\end{figure*}

\begin{figure*}
    {
   \subfigure[]{
     \includegraphics[width=9.2 cm]{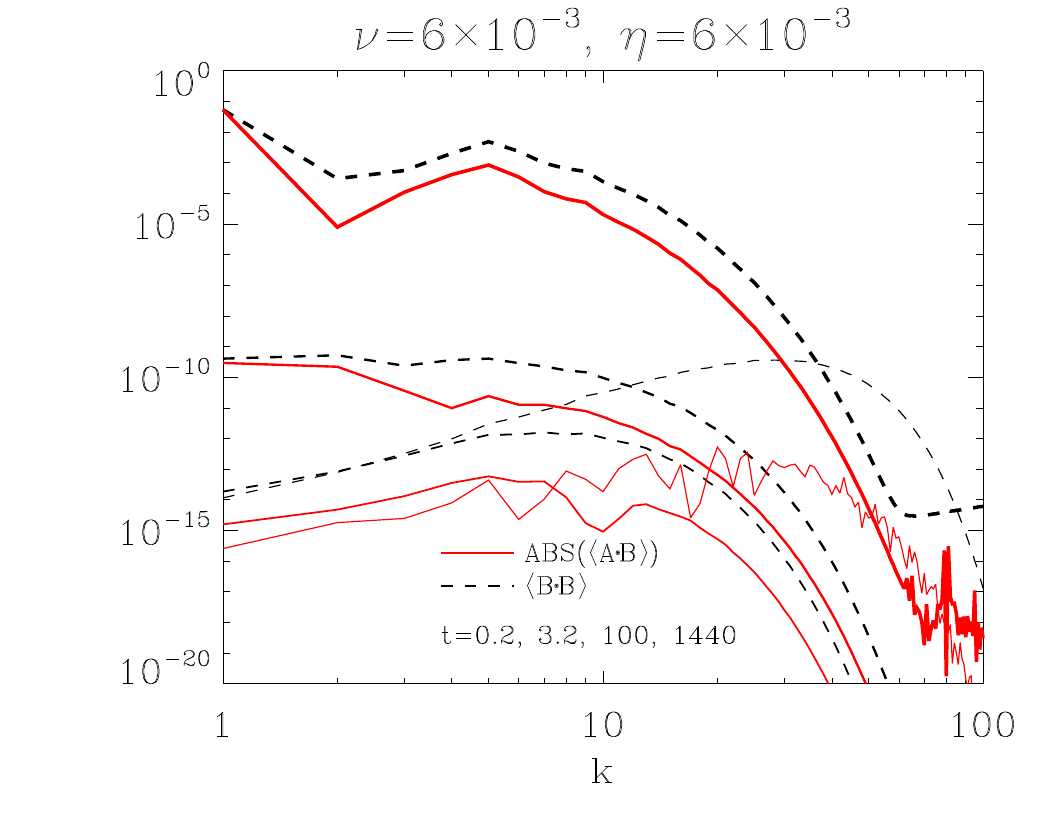}
     \label{f2a}
    }\hspace{-13 mm}
   \subfigure[]{
   \includegraphics[width=9.2 cm]{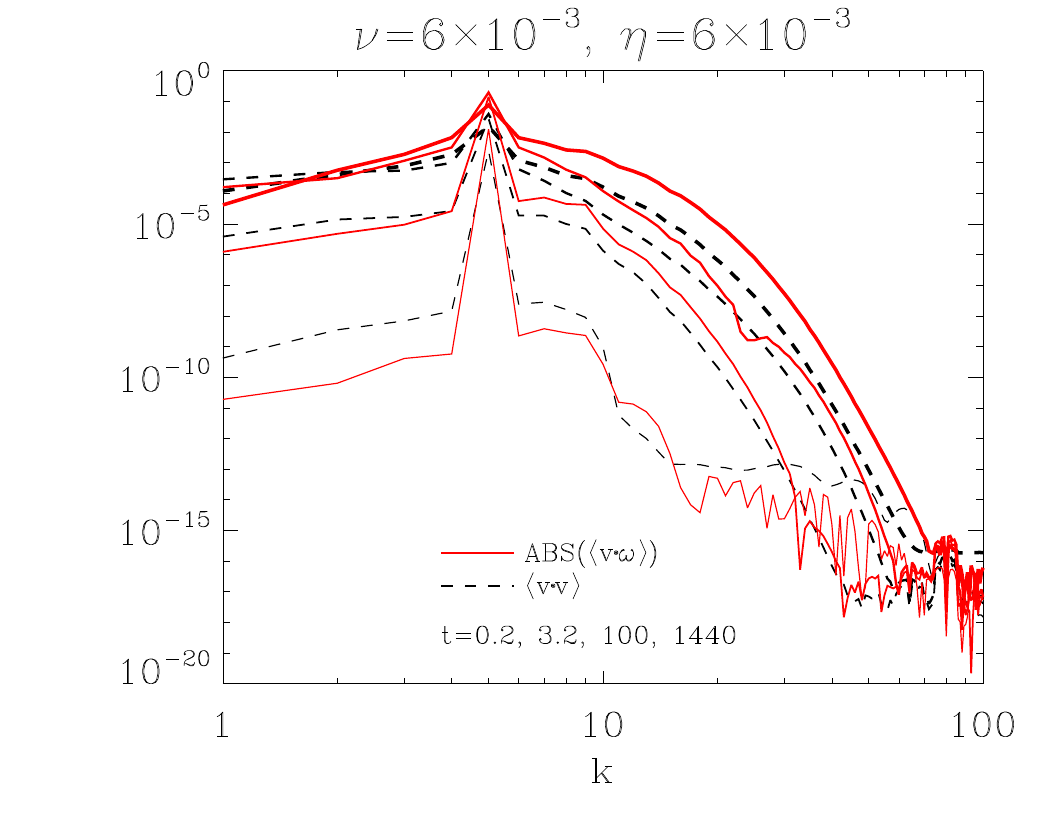}
     \label{f2b}
   }\hspace{-13 mm}
   \subfigure[]{
   \includegraphics[width=9.2 cm]{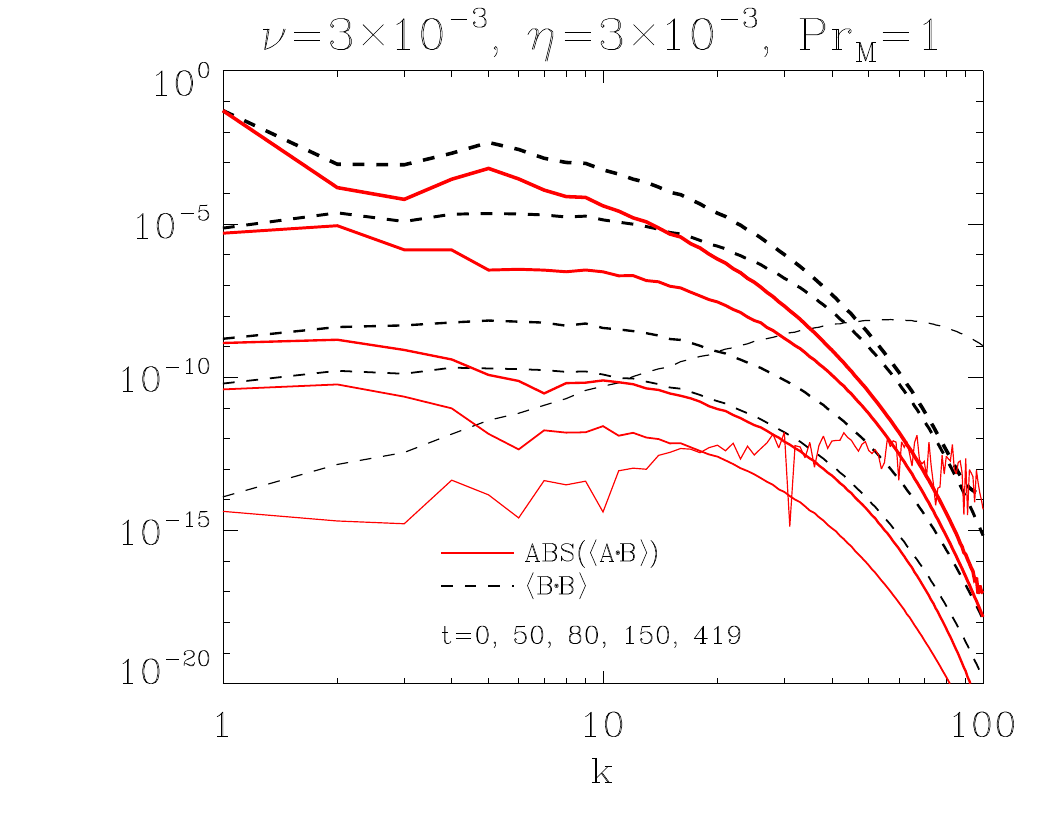}
     \label{f2c}
   }\hspace{-13 mm}
   \subfigure[]{
     \includegraphics[width=9.2 cm]{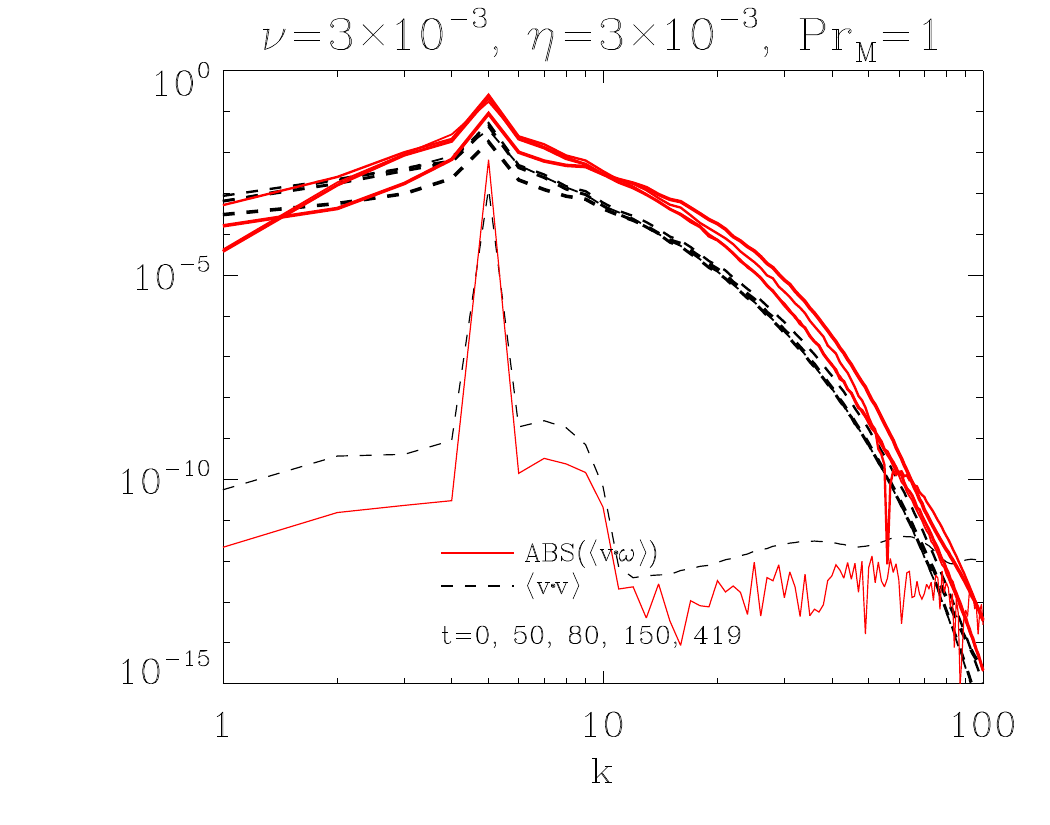}
     \label{f2d}
   }\hspace{-13 mm}
      \subfigure[]{
   \includegraphics[width=9.2 cm]{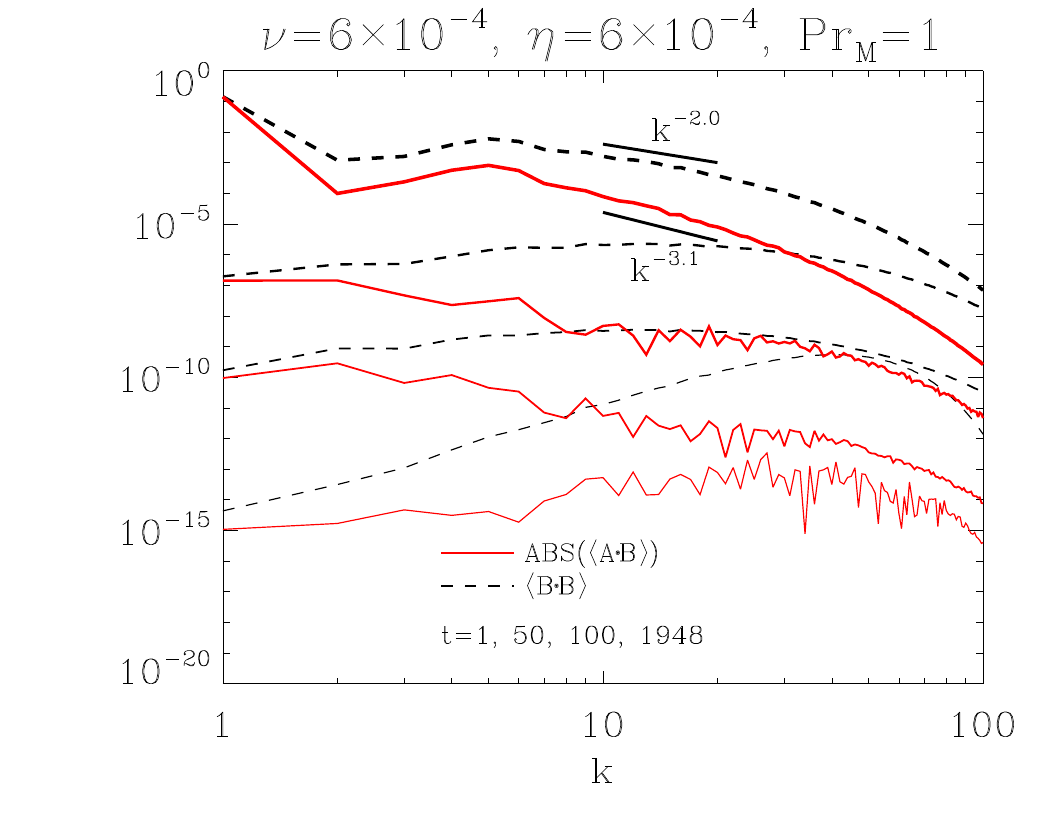}
     \label{f2e}
   }\hspace{-13 mm}
   \subfigure[]{
     \includegraphics[width=9.2 cm]{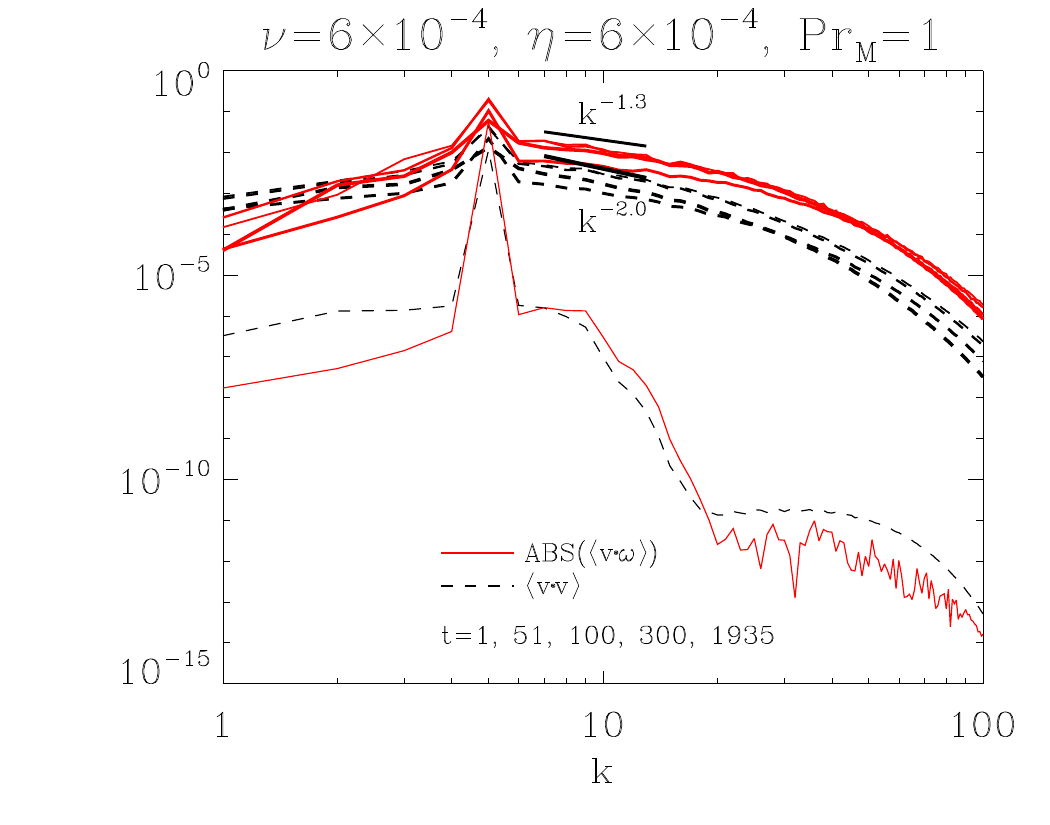}
     \label{f2f}
   }
}
\caption{Left panel: spectra of $|H_\mathrm{M}|$ and $2E_\mathrm{M}$, right panel : spectra of $|H_V|$ and $2E_V$. As $Re_M$ increases, the energy spectrum approaches Kolmogorov's law with a slope of $k^{-5/3}$ (approximately $k^{-1.67}$).}
\end{figure*}

\begin{figure*}
{
   \subfigure[]{
     \includegraphics[width=9.4 cm]{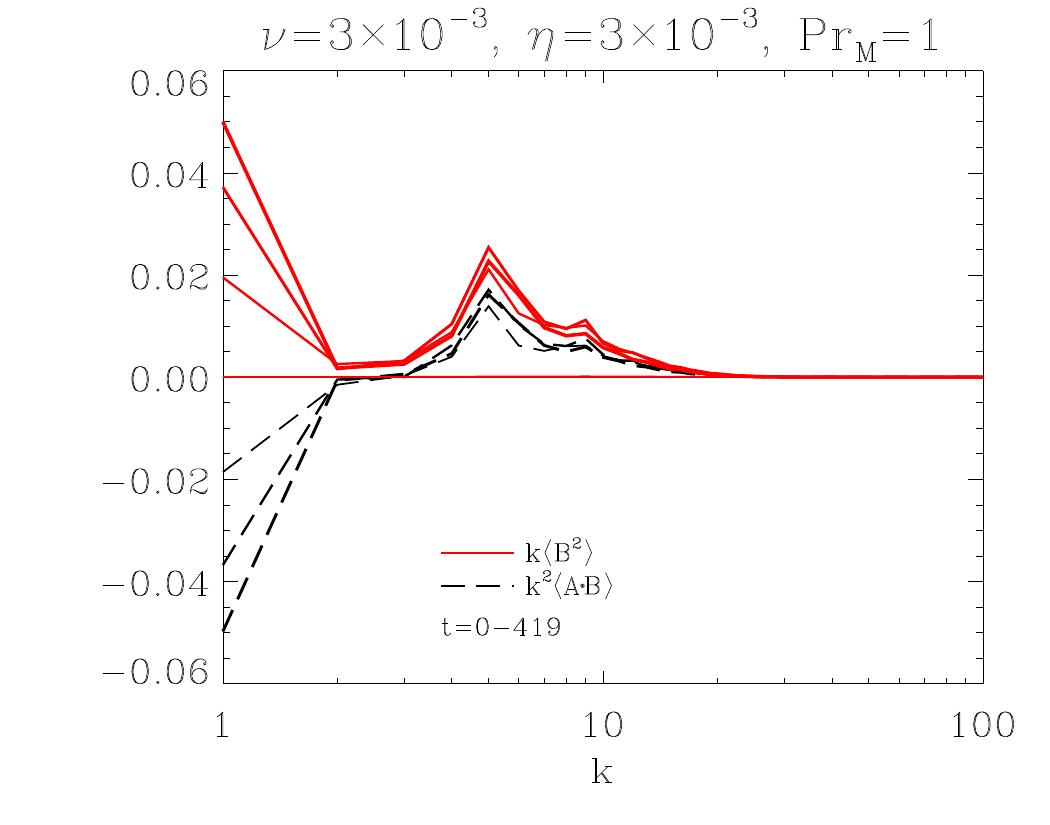}
     \label{f2a_new}
   }\hspace{-13 mm}
   \subfigure[]{
     \includegraphics[width=9.4 cm]{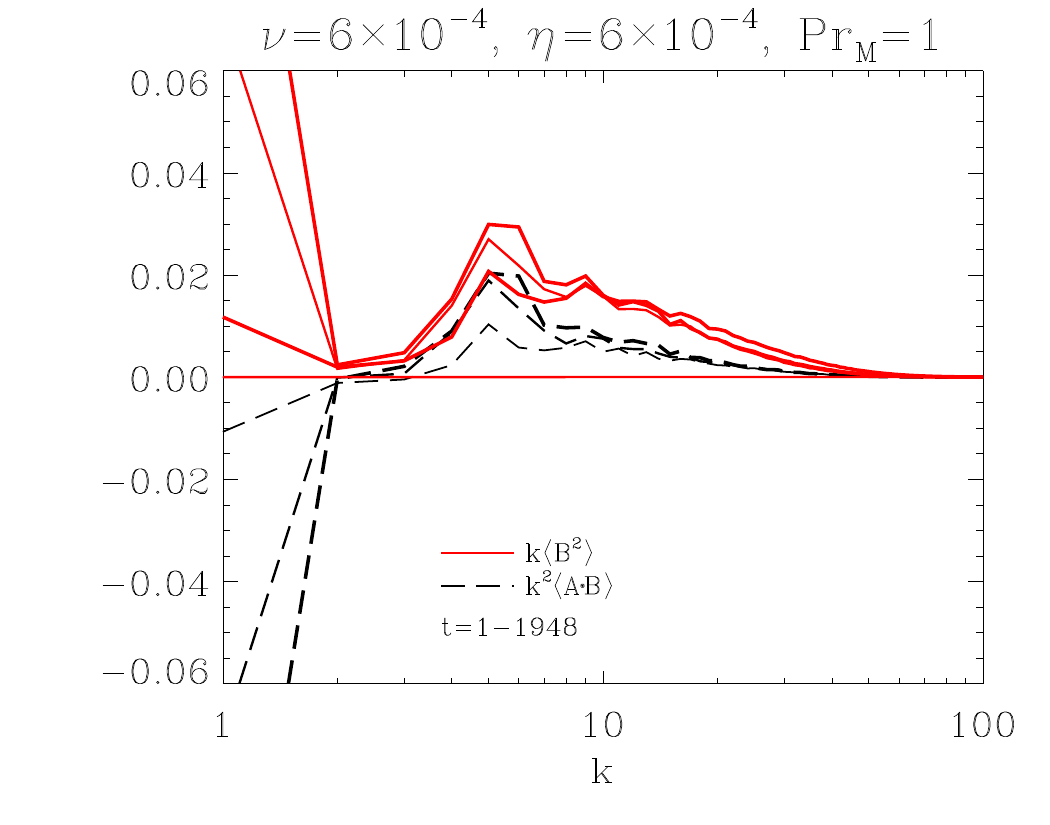}
     \label{f2b_new}
   }\hspace{-13 mm}
   \subfigure[]{
     \includegraphics[width=9.4 cm]{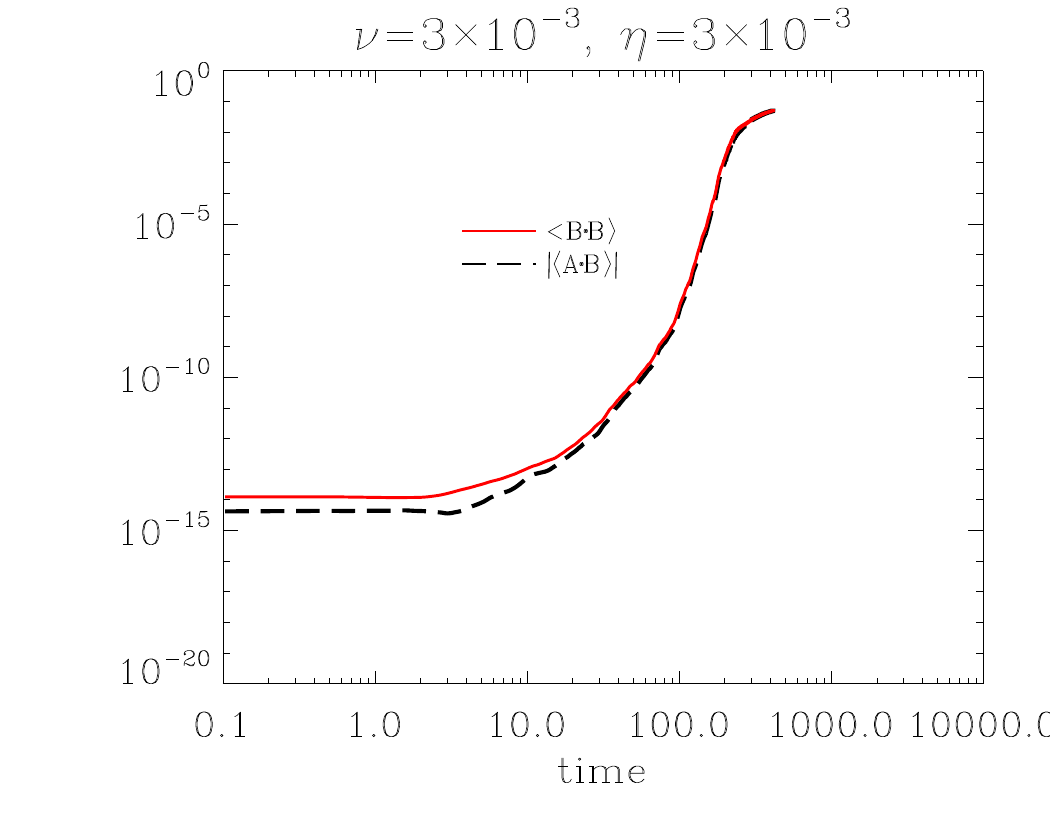}
     \label{f2c_new}
   }\hspace{-13 mm}
   \subfigure[]{
     \includegraphics[width=9.4 cm]{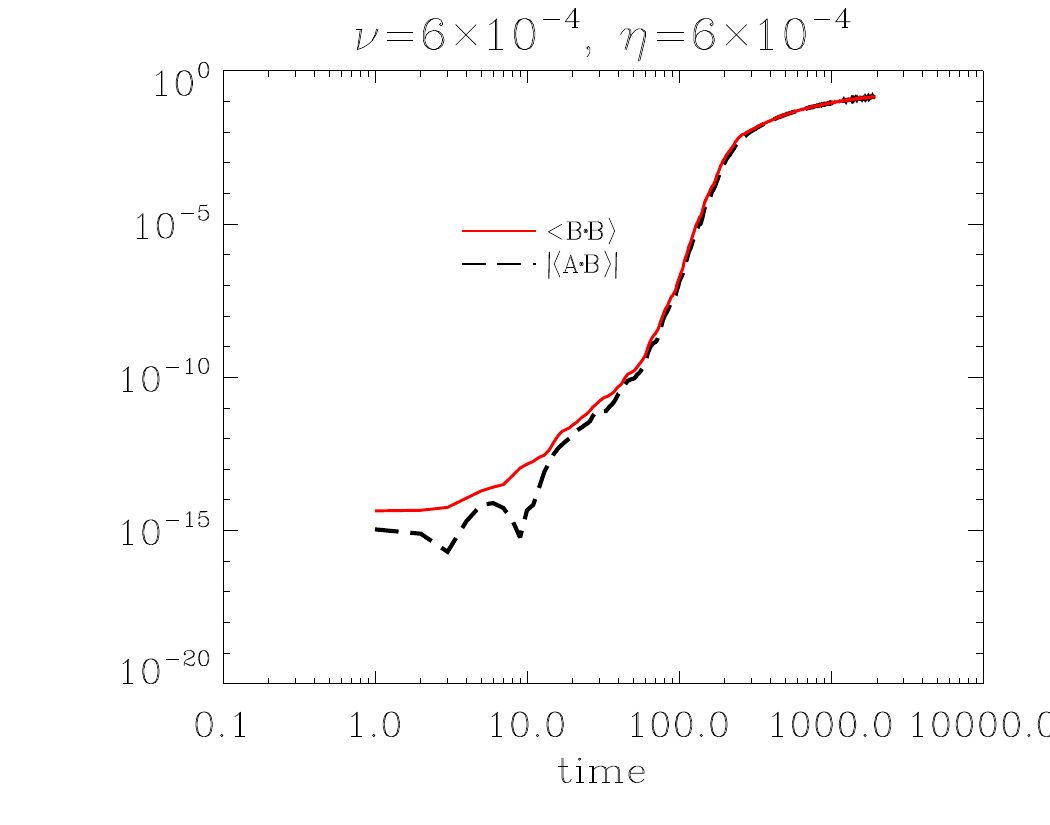}
     \label{f2d_new}
   }
}
\caption{
(a), (b) Current helicity $\langle \mathbf{J}\cdot \mathbf{B}\rangle \,(=k^2 H_\mathrm{M})$ and $k \langle B^2 \rangle$ in Fourier space. The polarity of $\langle \mathbf{J}\cdot \mathbf{B} \rangle$ and $H_\mathrm{M}$ at $k=1$ is opposite to that of magnetic helicity in small scales ($k>2$).
(c), (d) $\langle \overline{\mathbf{A}}\cdot \overline{\mathbf{B}}\rangle$ and $\langle \overline{\mathbf{B}^2} \rangle$ in real space. Their discrepancies decrease as the fields become saturated.
}
\end{figure*}

\begin{figure*}
{
   \subfigure[]{
     \includegraphics[width=7.4 cm]{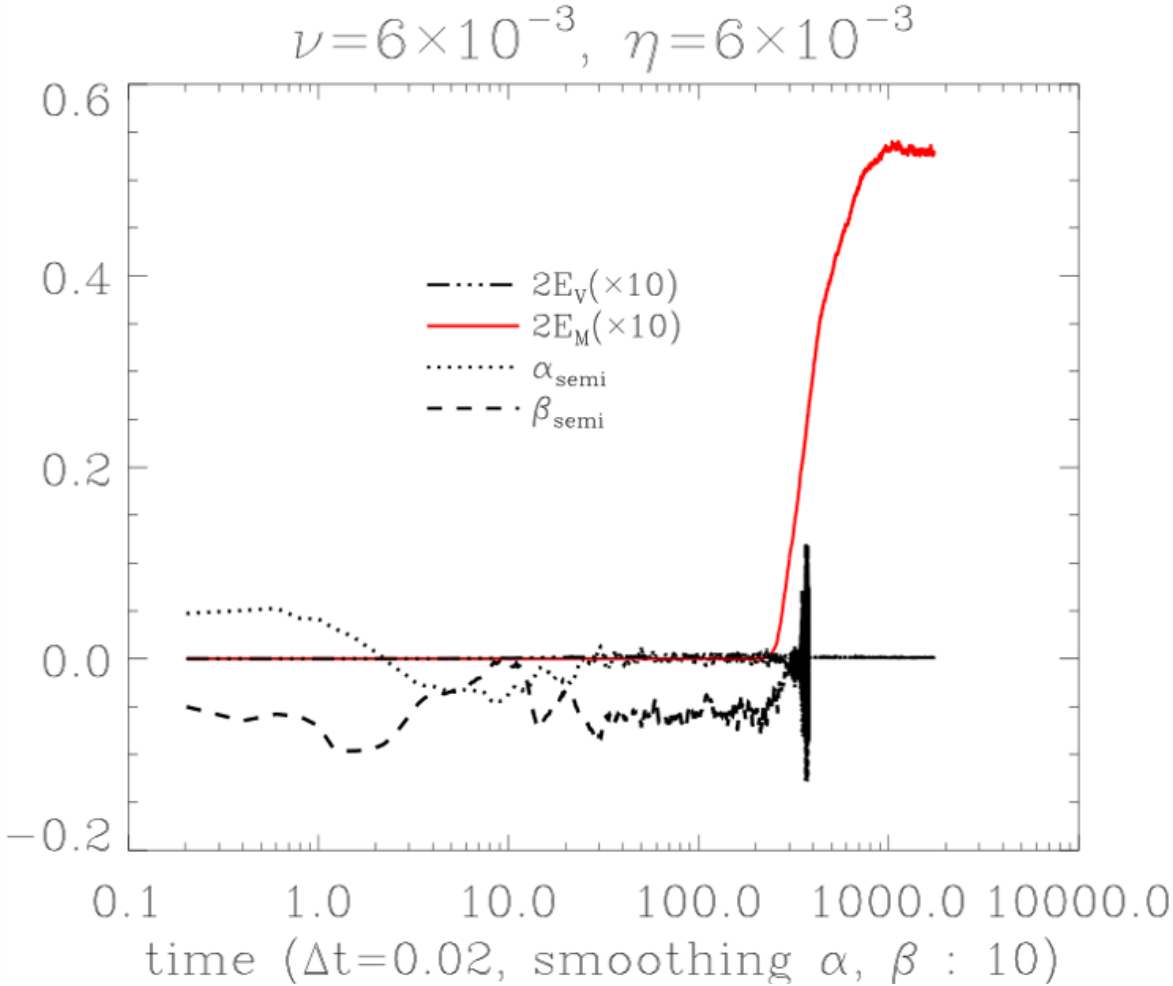}
     \label{f3a}
   }\hspace{-13 mm}
   \subfigure[]{
     \includegraphics[width=8.2 cm]{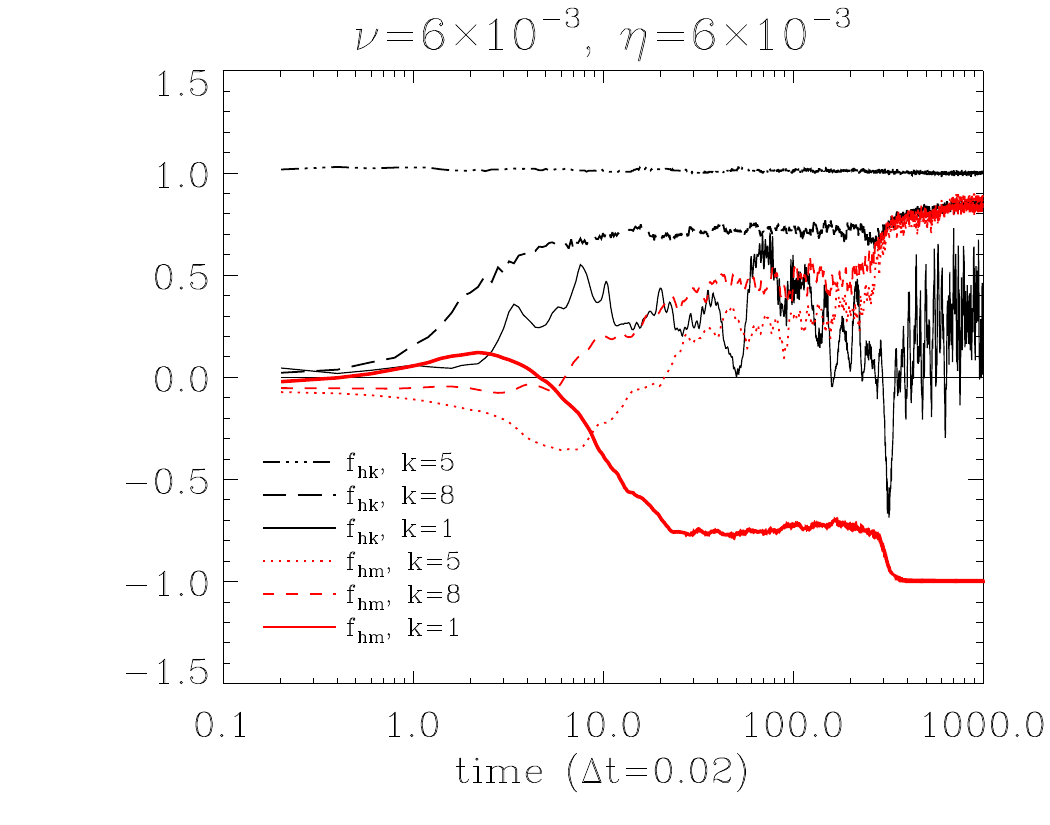}
     \label{f3b}
   }\hspace{-13 mm}
   \subfigure[]{
     \includegraphics[width=8.2 cm]{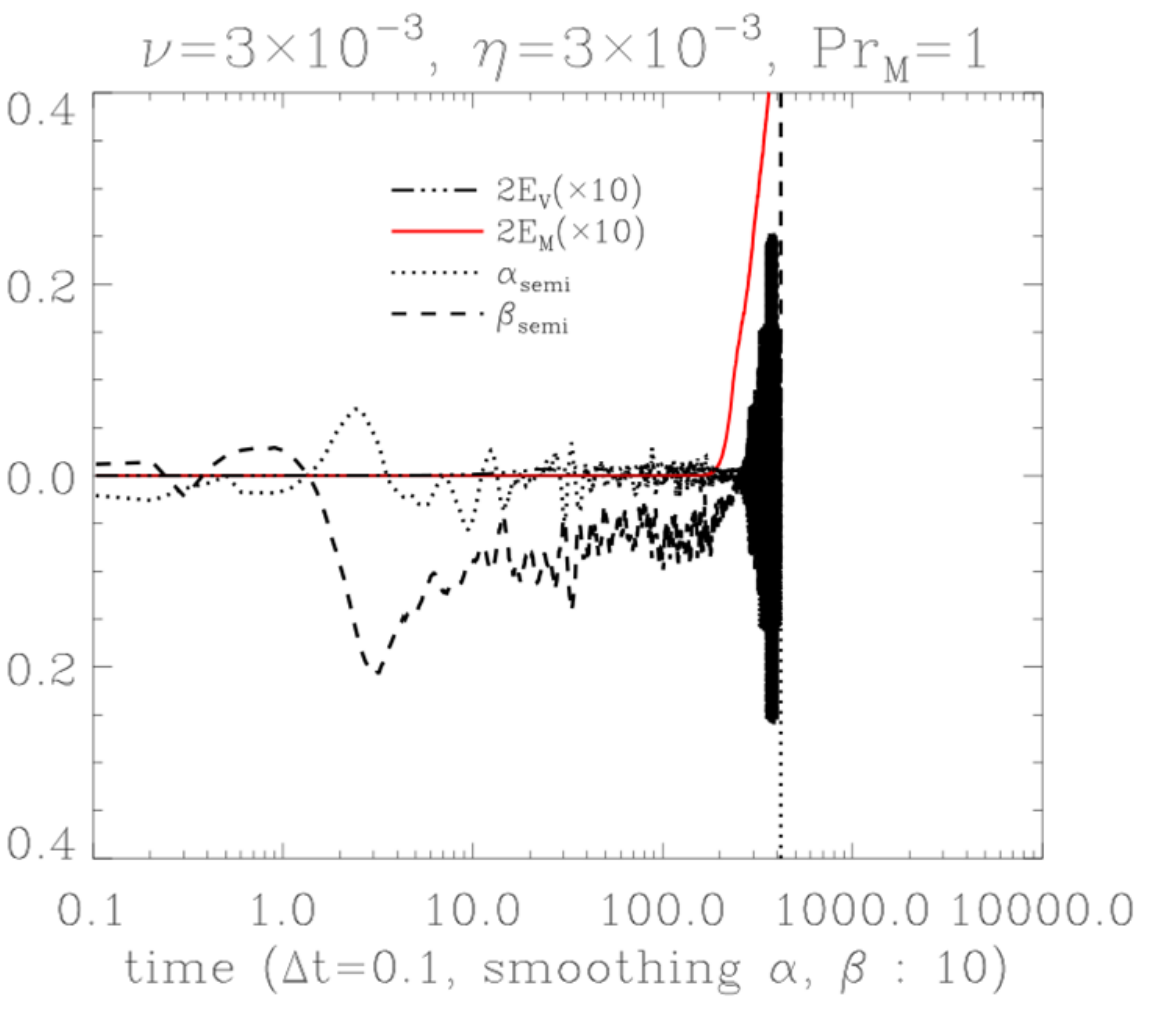}
     \label{f3c}
   }\hspace{-13 mm}
   \subfigure[]{
     \includegraphics[width=9.2 cm]{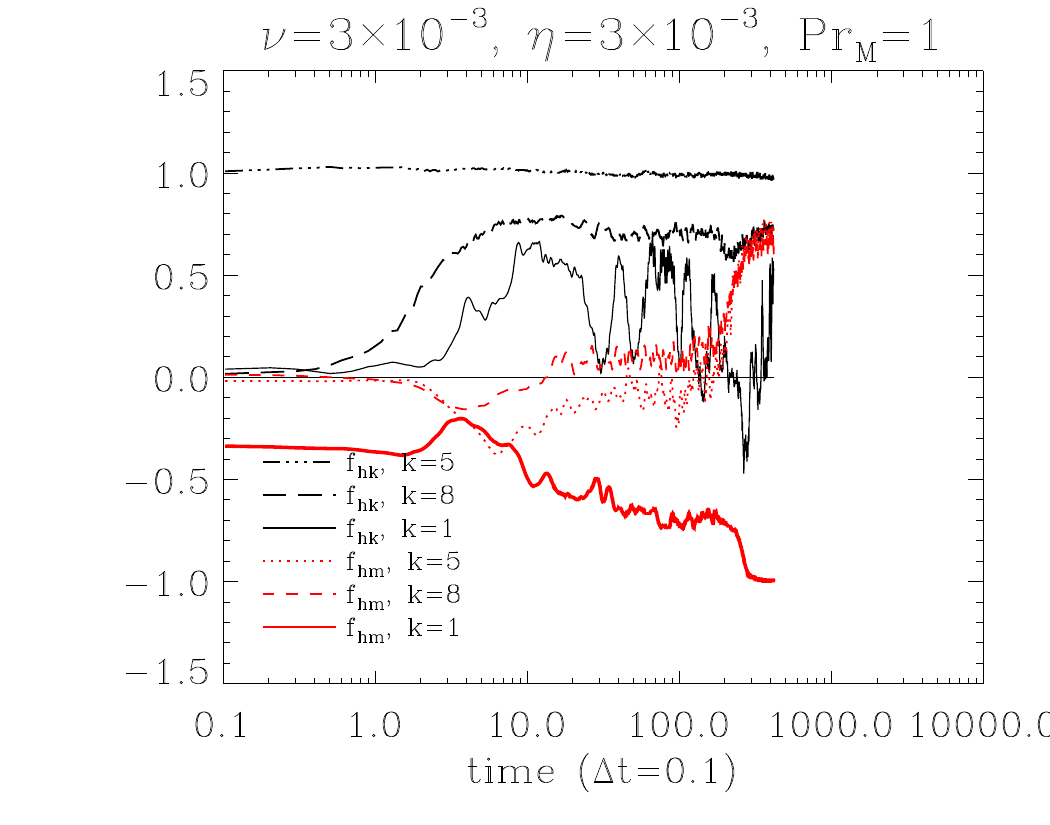}
     \label{f3d}
   }\hspace{-13 mm}
   \subfigure[]{
     \includegraphics[width=8.2 cm]{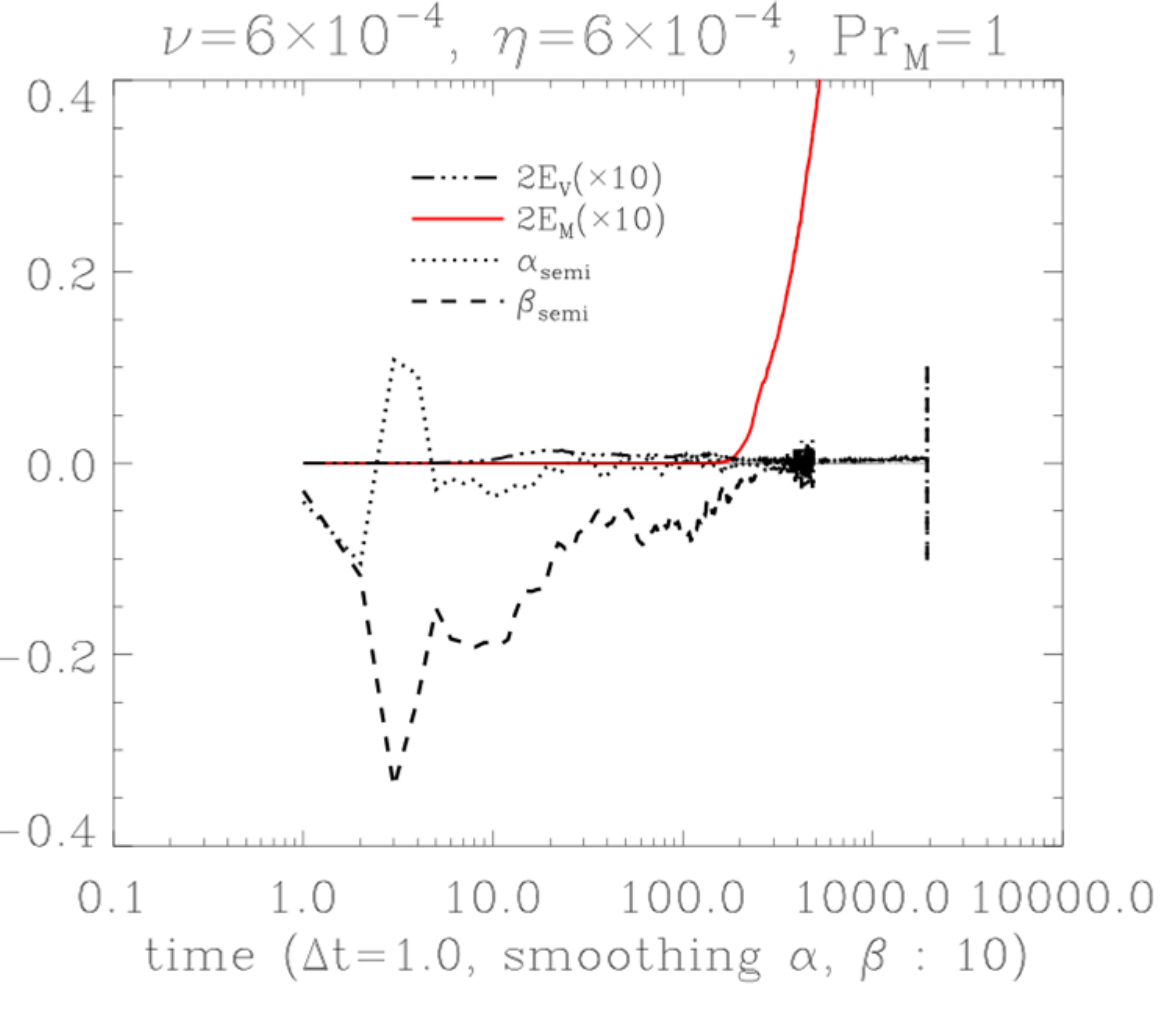}
     \label{f3e}
   }\hspace{-13 mm}
   \subfigure[]{
     \includegraphics[width=9.2 cm]{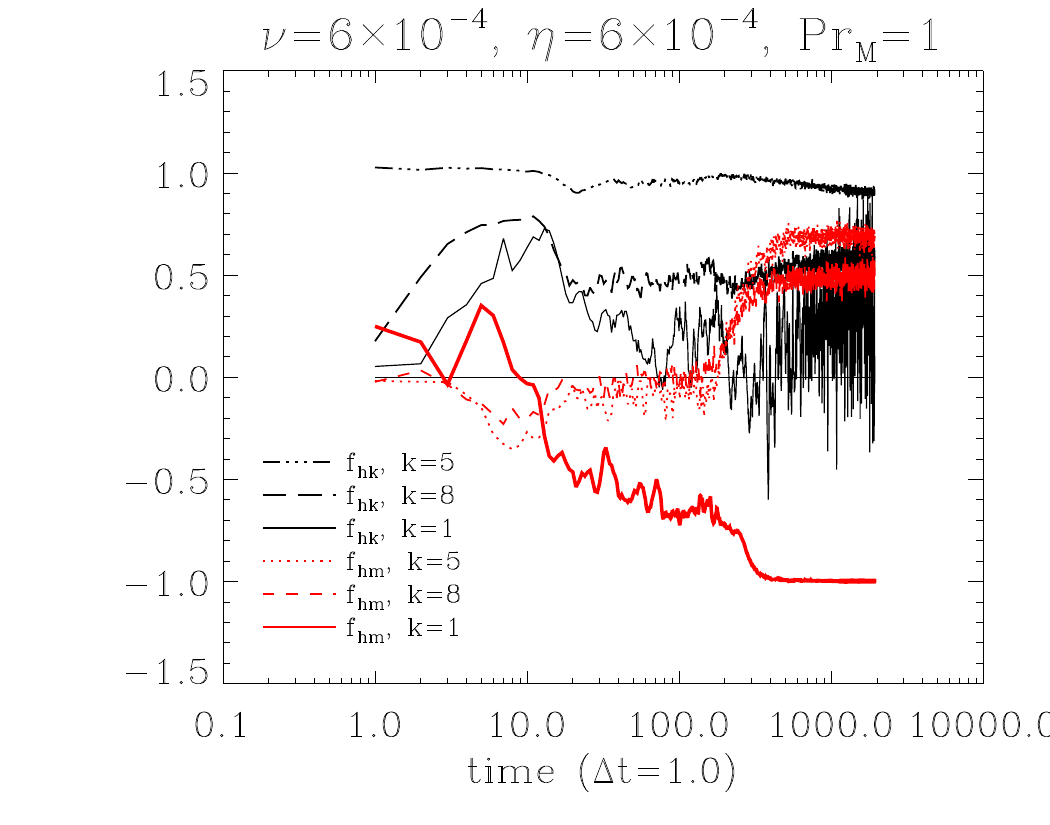}
     \label{f3f}
   }
}
\caption{
Left panels: $\alpha$ and $\beta$ are computed from $\overline{H}_\mathrm{M}(t)$ and $\overline{E}_\mathrm{M}(t)$. $\alpha$ and $\beta$ profiles represent averaged values over 10 adjacent points.
Right panels: kinetic helicity ratio $f_\mathrm{hk} = \langle \mathbf{U} \cdot (\nabla \times \mathbf{U}) \rangle / k \langle U^2 \rangle$ and magnetic helicity ratio $f_\mathrm{hm} = k \langle \mathbf{A} \cdot \mathbf{B} \rangle / \langle B^2 \rangle$.
Panel (a): simulation output recorded at time intervals $\Delta t = 0.02$, $Re_M = 261$.
Panel (b): $f_\mathrm{hk}$ and $f_\mathrm{hm}$ at $k = 1$ (large scale), $k = 5$ (forcing scale), and $k = 8$ (small scale), $Re_M = 261$.
Panels (c), (d): $Re_M = 607$.
Panels (e), (f): $Re_M = 4290$.
}
\end{figure*}

\begin{figure*}
{
   \subfigure[]{
     \includegraphics[width=8.5 cm]{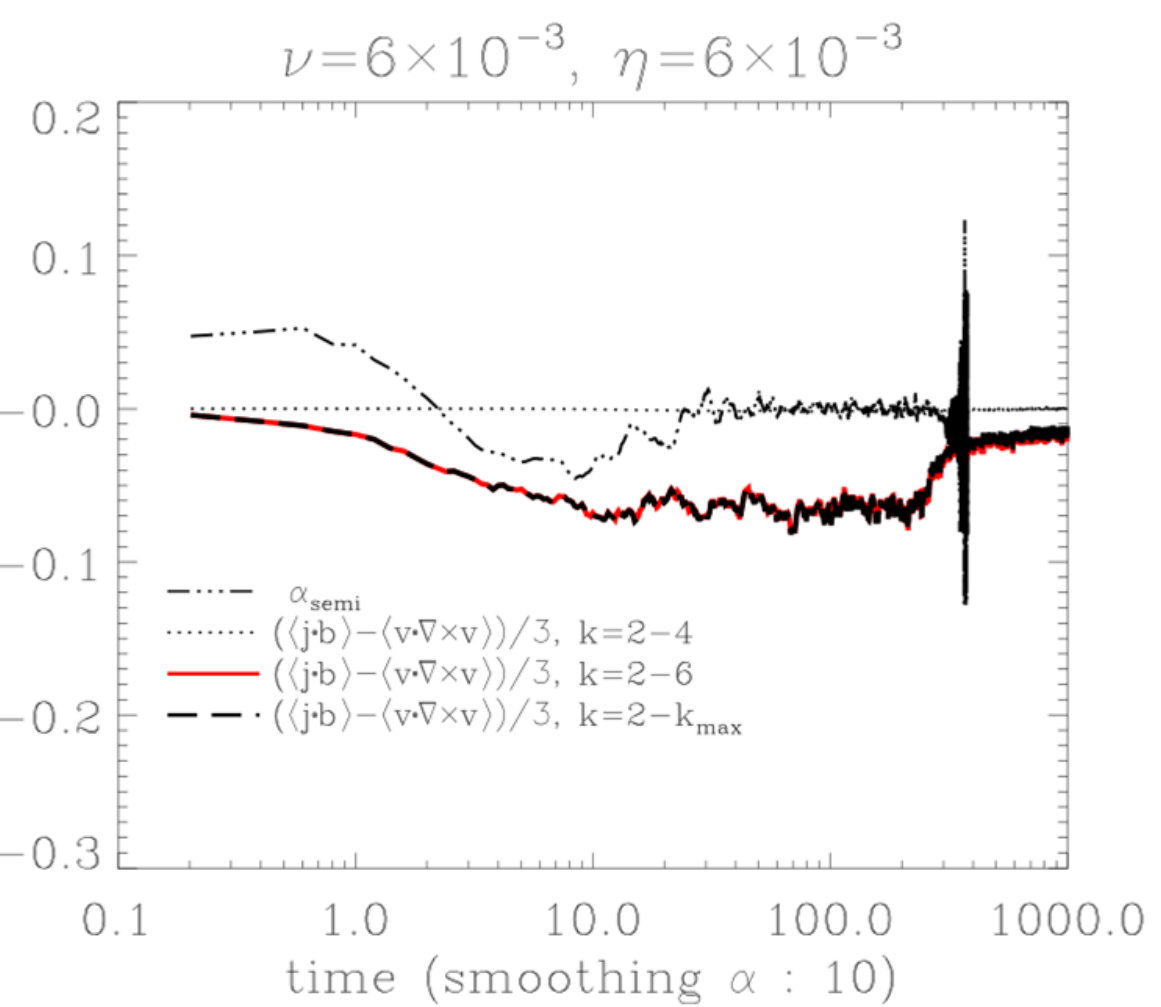}
     \label{f4a}
   }\hspace{-6 mm}
   \subfigure[]{
     \includegraphics[width=8.5 cm]{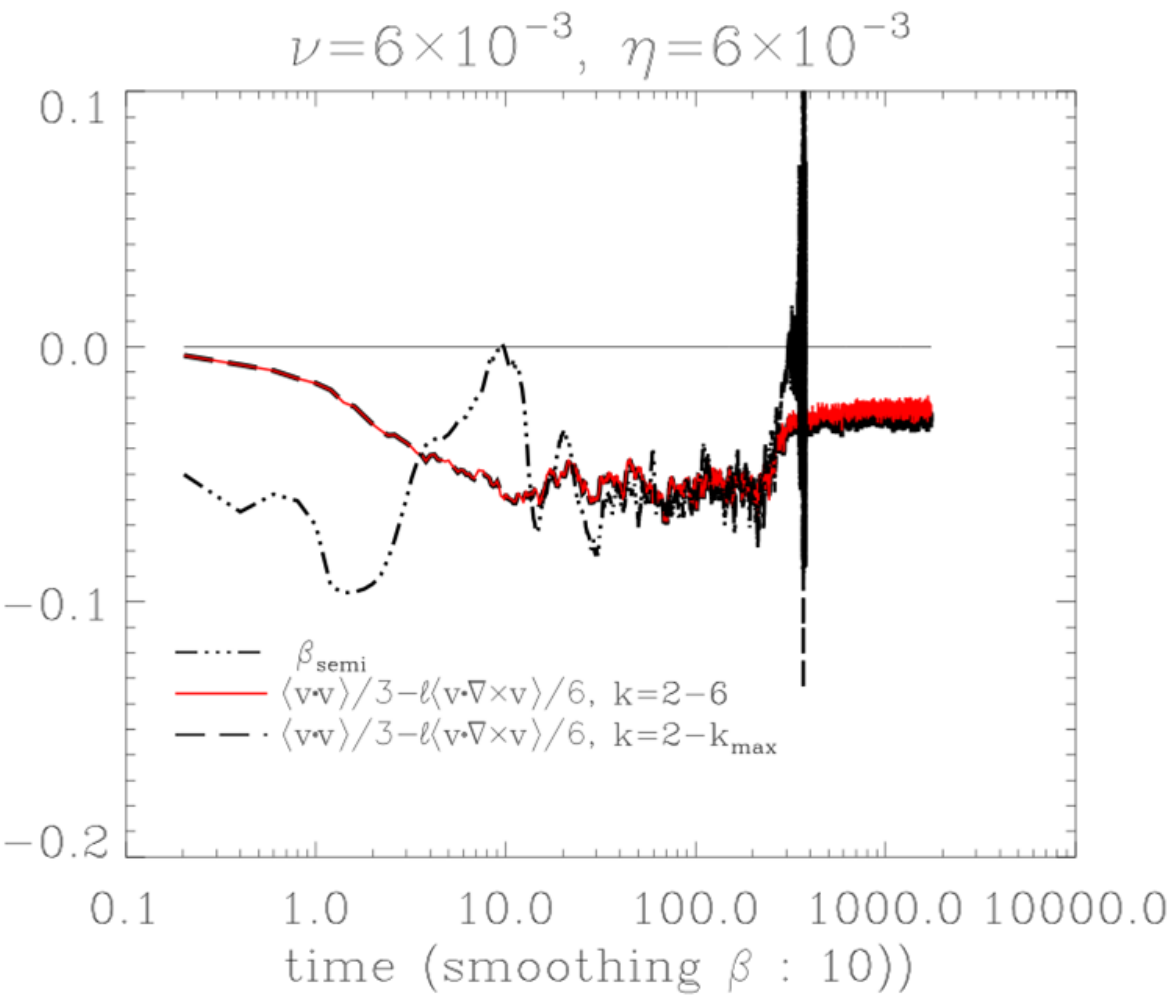}
     \label{f4b}
   }\hspace{-6 mm}
   \subfigure[]{
     \includegraphics[width=8.5 cm]{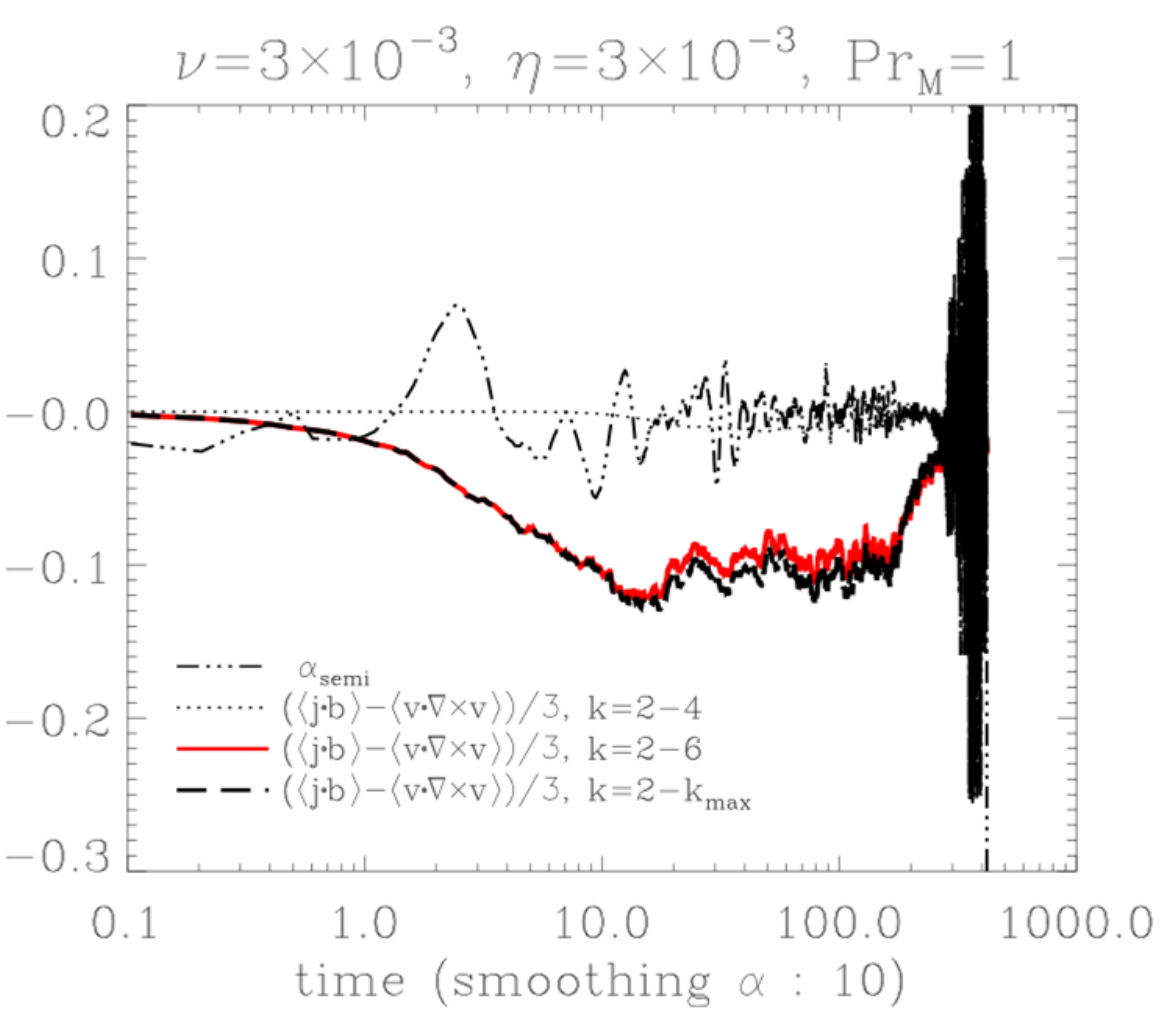}
     \label{f4c}
   }\hspace{-6 mm}
   \subfigure[]{
     \includegraphics[width=8.5 cm]{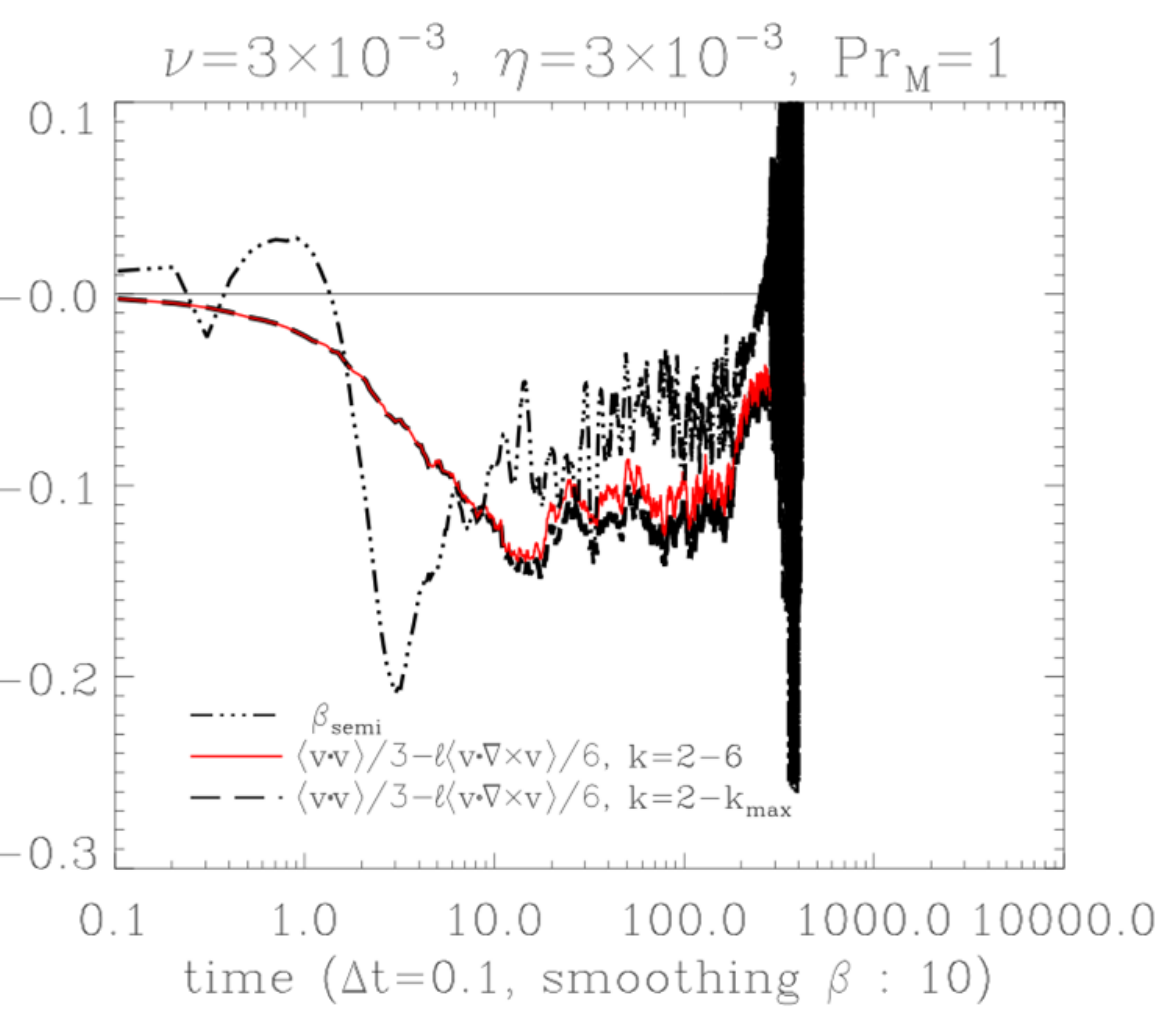}
     \label{f4d}
   }\hspace{-6 mm}
   \subfigure[]{
     \includegraphics[width=8.5 cm]{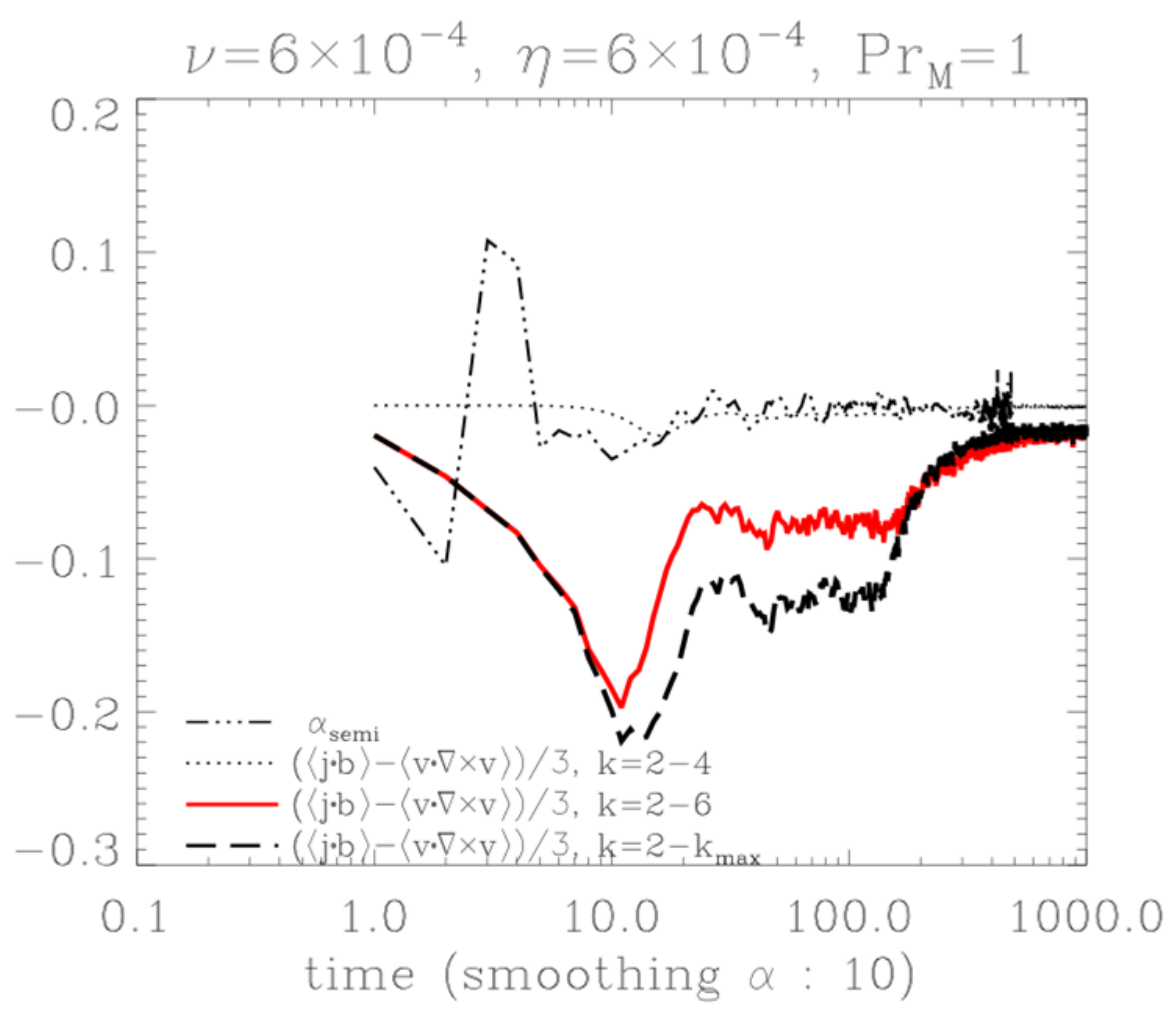}
     \label{f4e}
   }\hspace{-6 mm}
   \subfigure[]{
     \includegraphics[width=8.5 cm]{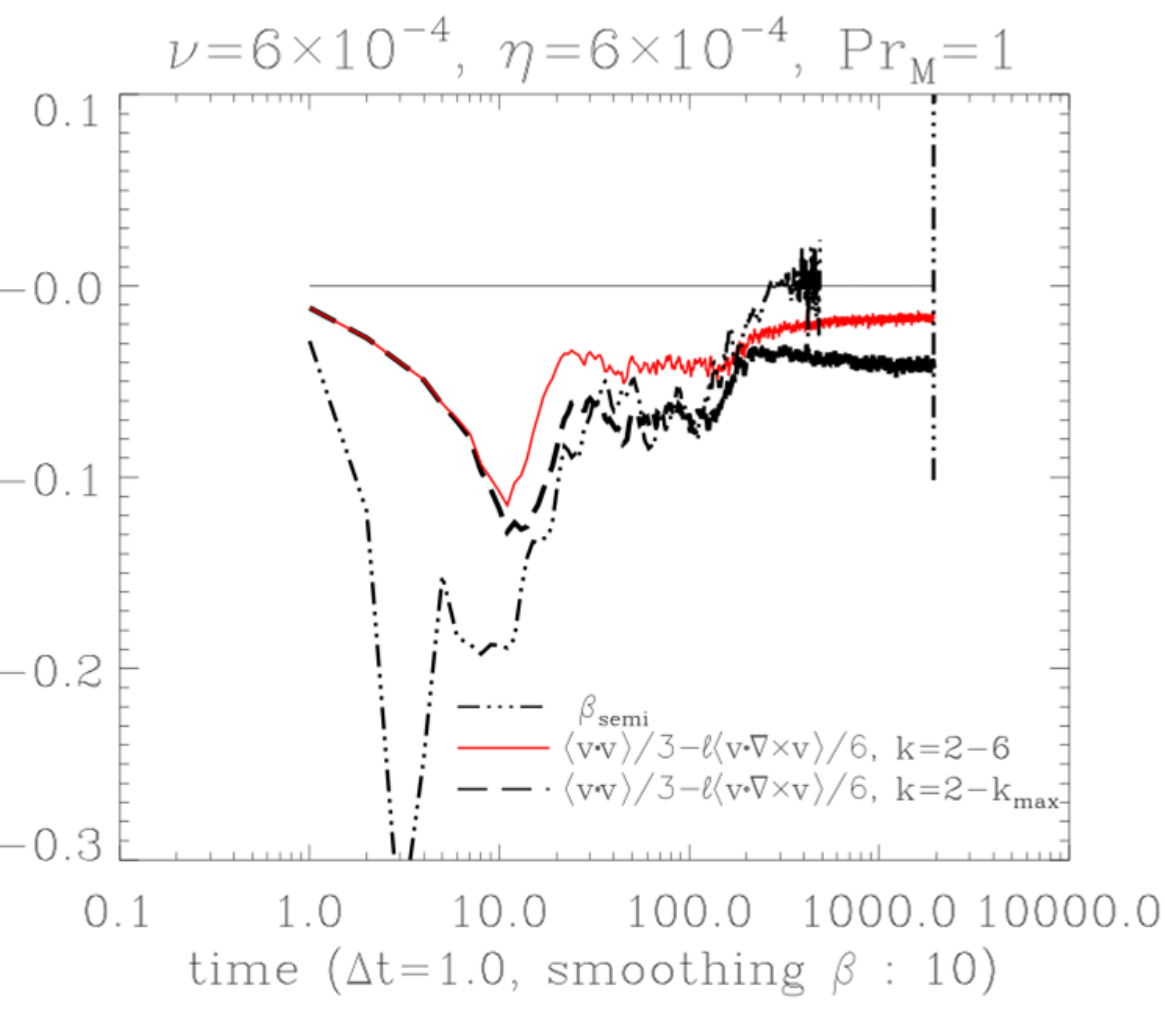}
     \label{f4f}
   }
}
\caption{
The profiles of $\alpha$ and $\beta$ at different magnetic Reynolds numbers, $Re_M = 261$, $607$, and $4290$.
Panel (a): $\alpha_{semi}$ from Eq.~(\ref{alphaSolution3}) and $\alpha_{MFT}$ from Eq.~(\ref{alpha_beta_MFT}) for $Re_M = 261$.
Panel (b): $\beta_{semi}$ from Eq.~(\ref{betaSolution3}) and $\beta_{theo}$ from Eq.~(\ref{general_beta_derivation8}) for $Re_M = 261$.
Panels (c), (d): $\alpha$ and $\beta$ profiles for $Re_M = 607$.
Panels (e), (f): $\alpha$ and $\beta$ profiles for $Re_M = 4290$.
}
\end{figure*}

\begin{figure*}
{
   \subfigure[]{
     \includegraphics[width=6.7 cm]{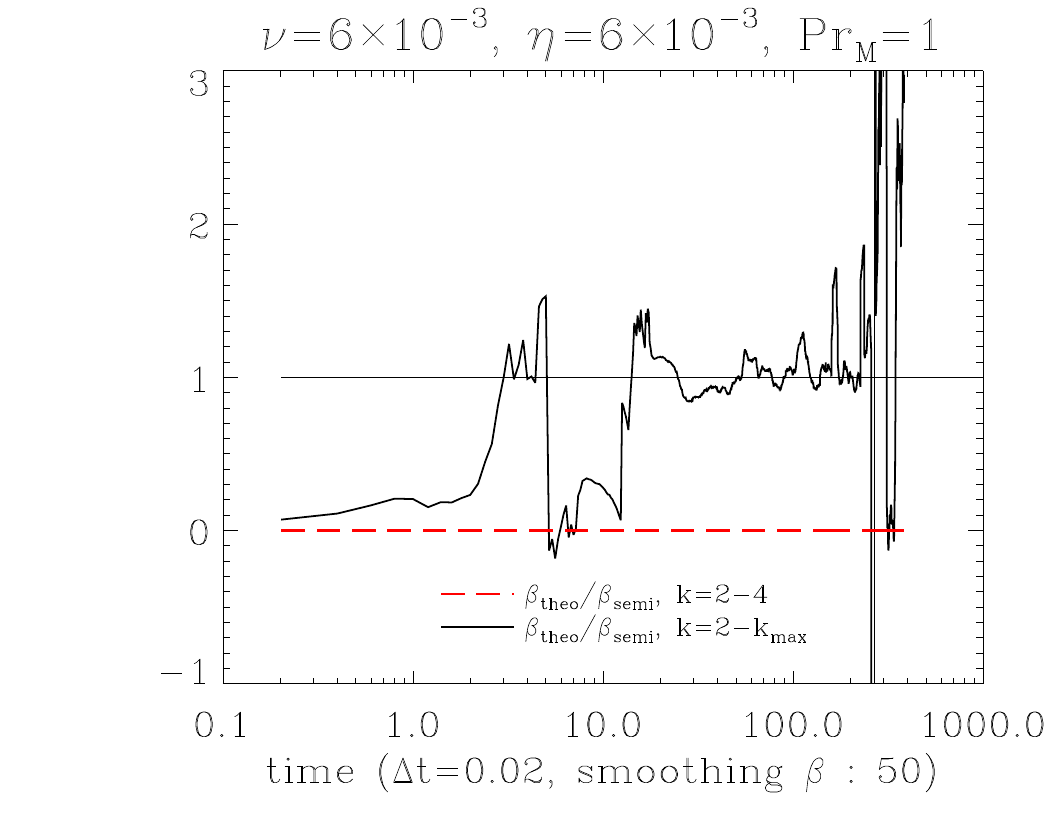}
     \label{f4a_new}
   }\hspace{-15 mm}
   \subfigure[]{
     \includegraphics[width=6.7 cm]{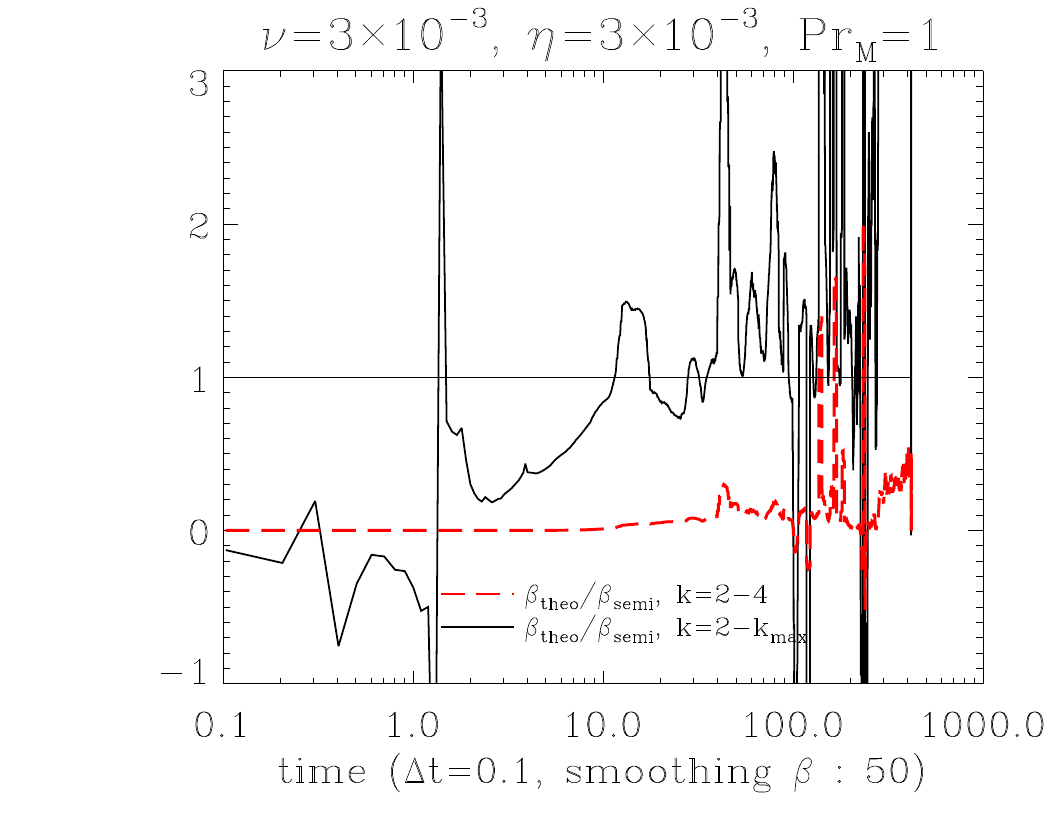}
     \label{f4b_new}
   }\hspace{-15 mm}
   \subfigure[]{
     \includegraphics[width=6.7 cm]{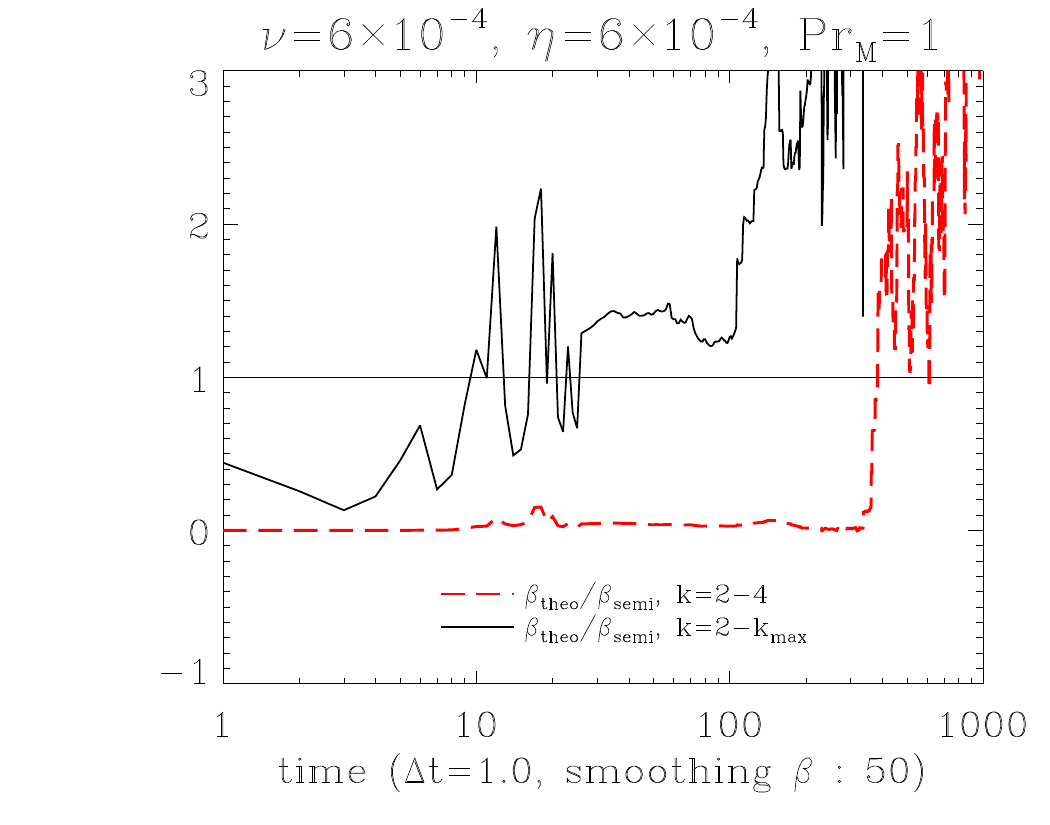}
     \label{f4c_new}
   }
}
\caption{
The ratio $\beta_{\mathrm{theo}}/\beta_{\mathrm{semi}}$ at different conditions.
}
\end{figure*}

\begin{figure*}
{
   \subfigure[]{
     \includegraphics[width=8.8 cm]{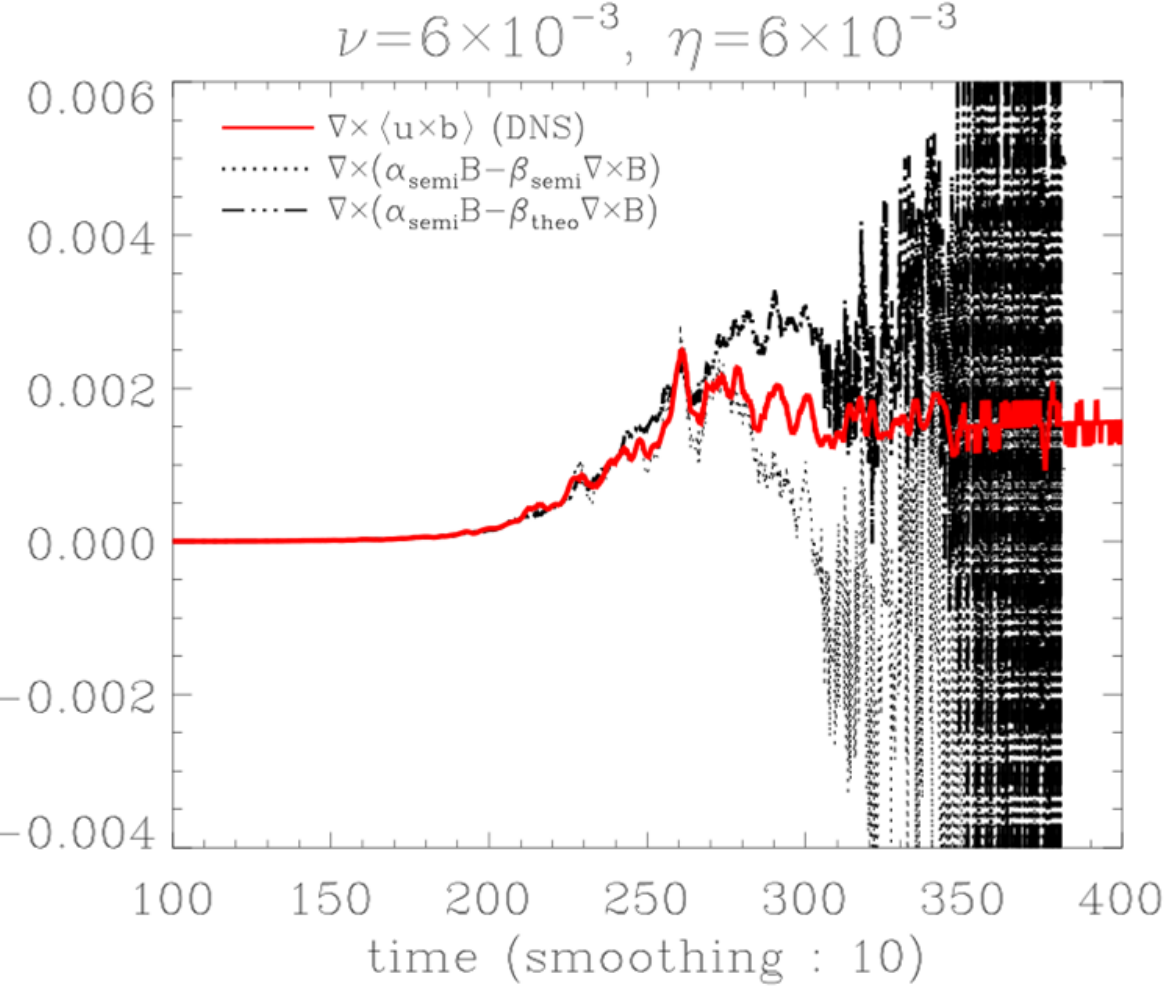}
     \label{f5a}
   }\hspace{-13 mm}
   \subfigure[]{
     \includegraphics[width=8.6 cm]{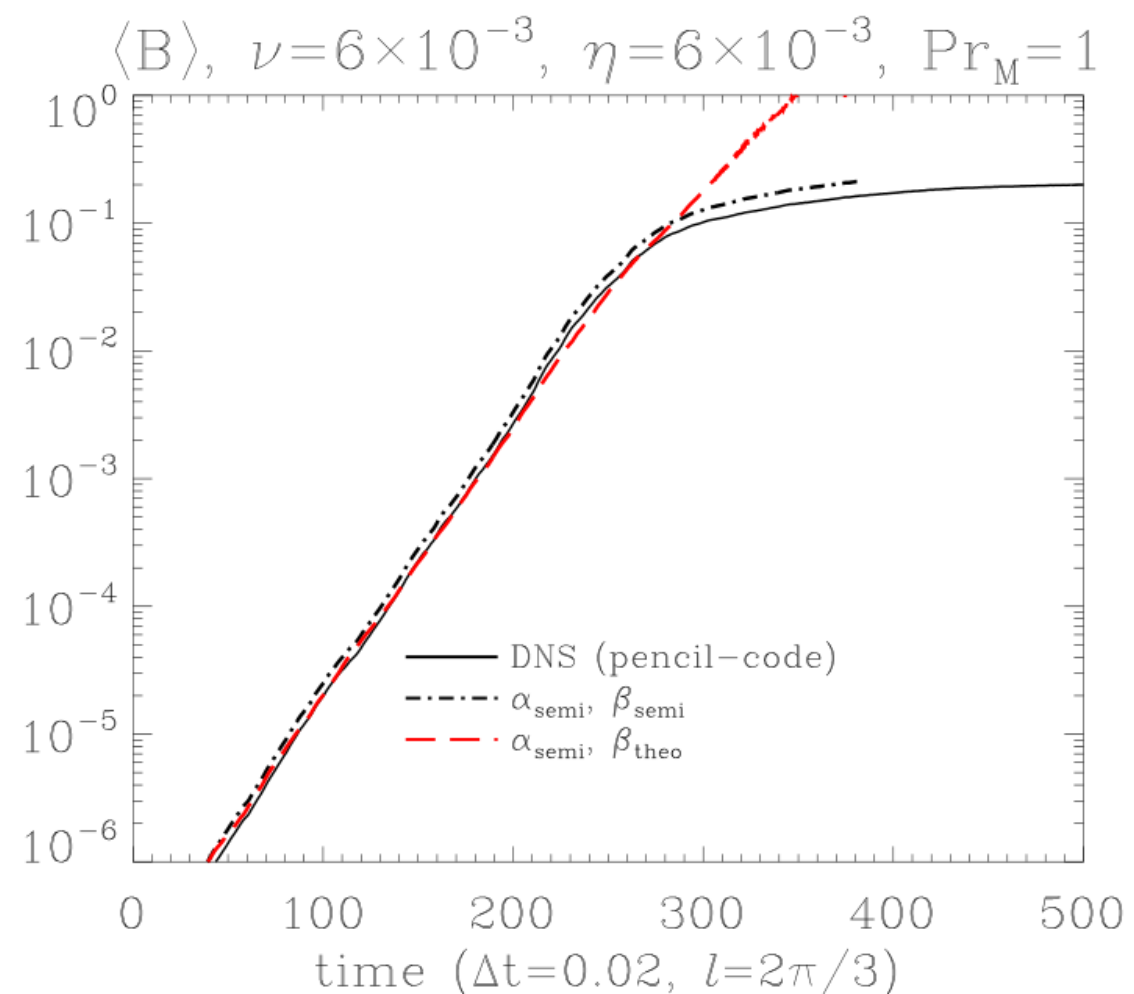}
     \label{f5b}
   }\hspace{-13 mm}
   \subfigure[]{
     \includegraphics[width=8.4 cm]{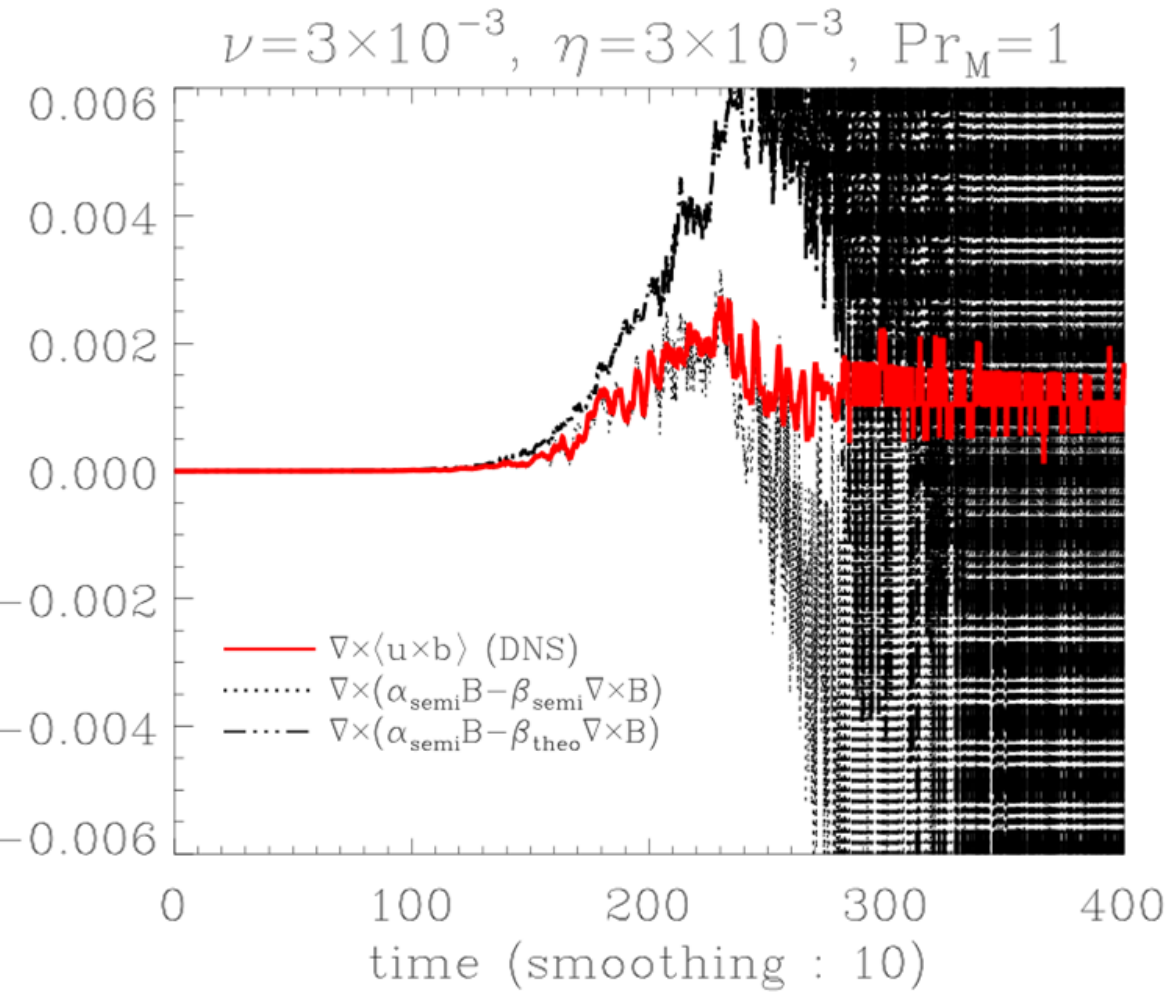}
     \label{f5c}
   }\hspace{-13 mm}
   \subfigure[]{
     \includegraphics[width=9.2 cm]{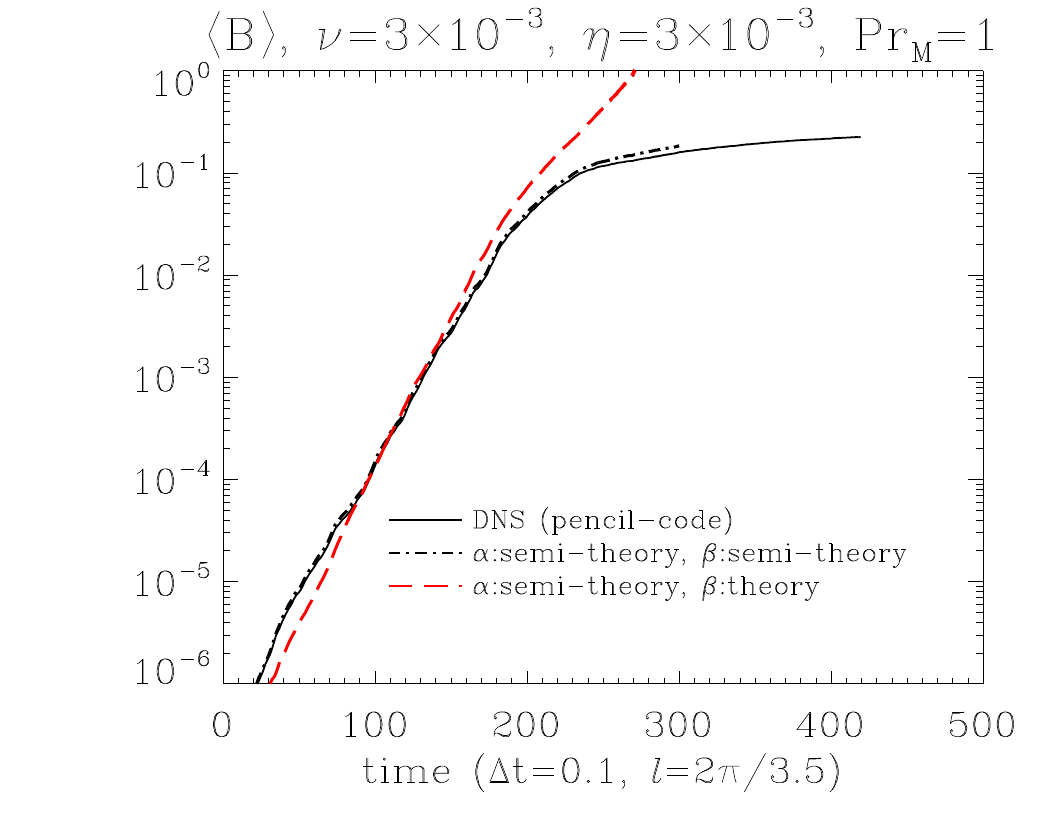}
     \label{f5d}
   }\hspace{-13 mm}
   \subfigure[]{
     \includegraphics[width=8.8 cm]{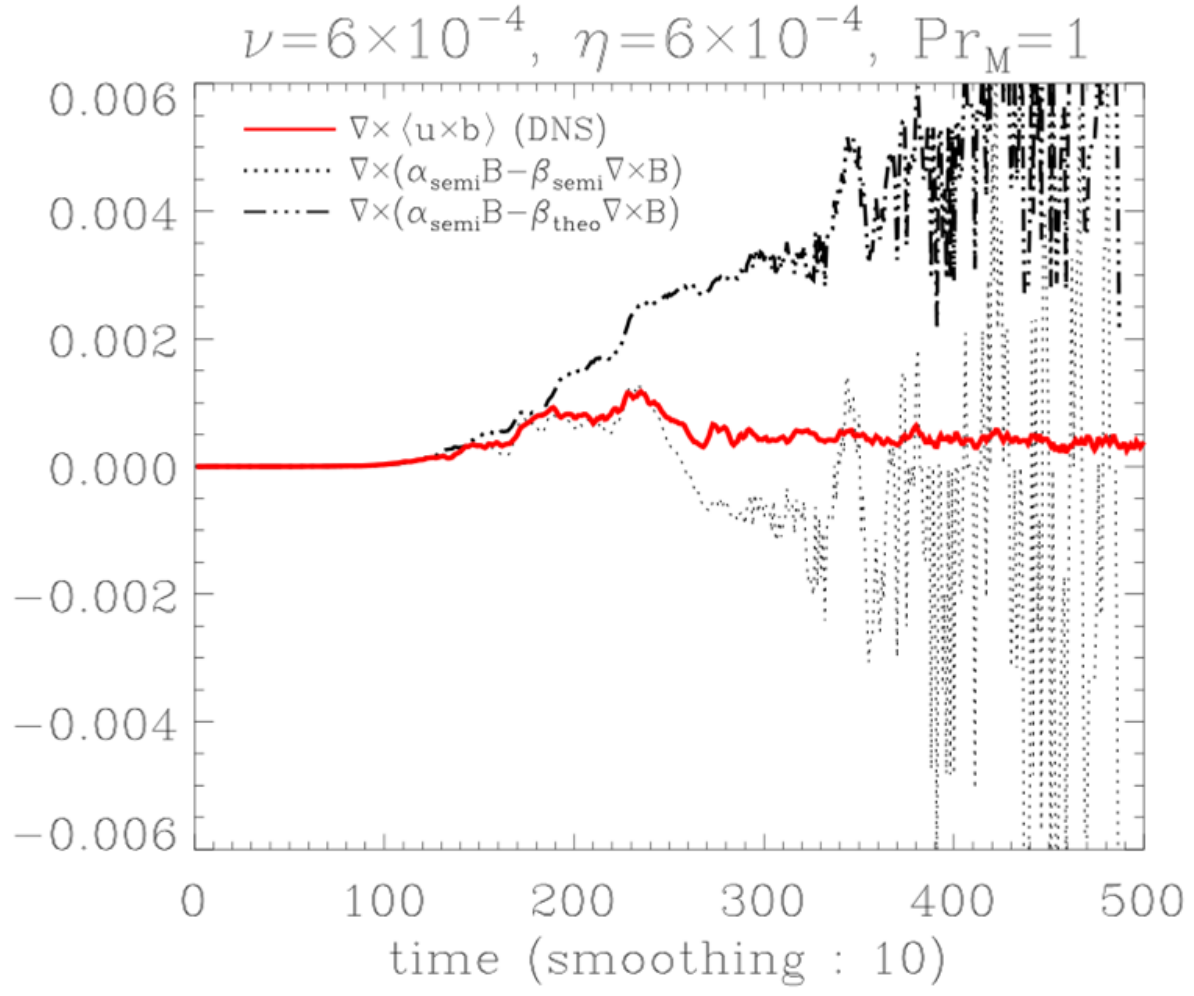}
     \label{f5e}
   }\hspace{-13 mm}
   \subfigure[]{
     \includegraphics[width=9.4 cm]{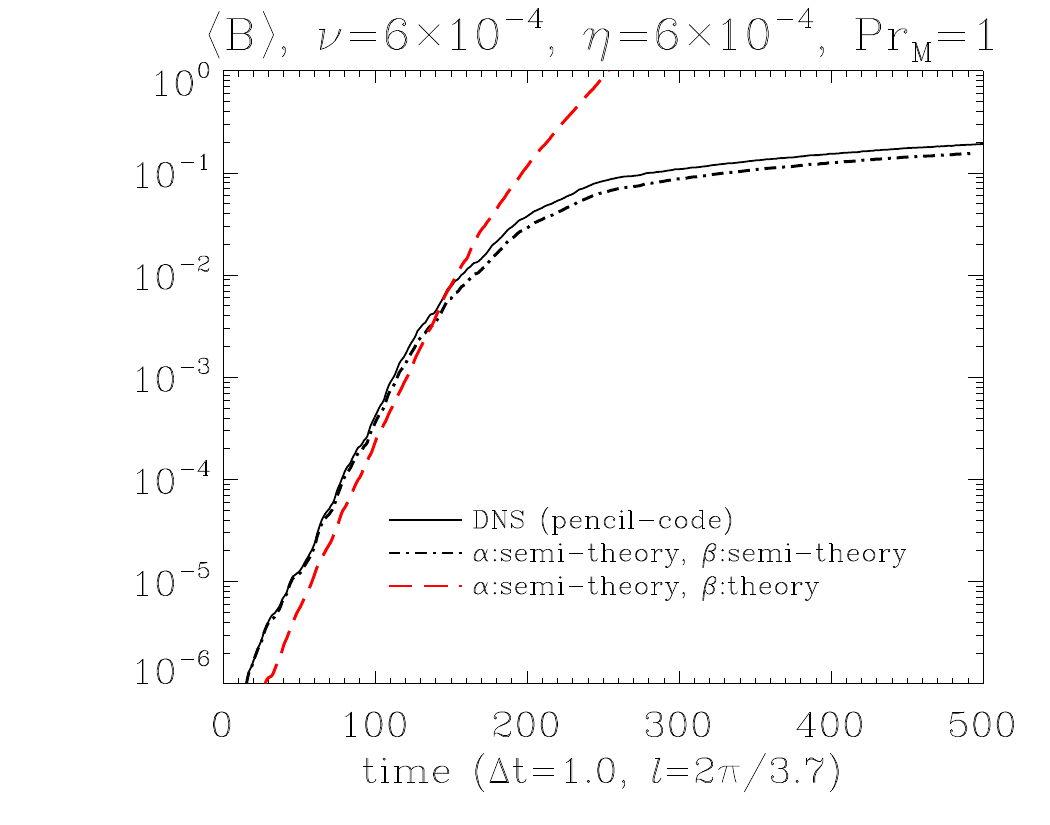}
     \label{f5f}
   }
}
\caption{
Left panels: $\nabla \times \mathrm{EMF}$.
Right panels: mean magnetic field $\overline{B}(t)$.
Here, $\alpha_{\mathrm{semi}}$ and $\beta_{\mathrm{semi}}$ are obtained from large-scale magnetic quantities, $\overline{H}_\mathrm{M}(t)$ and $\overline{E}_\mathrm{M}(t)$.
$\beta_{\mathrm{theo}}$ is calculated using turbulent kinetic quantities, $H_V$ and $E_V$ in the range $k = 2$ to $k_{\mathrm{max}}$.
}
\end{figure*}

\section{Methods}

\subsection{Numerical Method}
For our numerical simulations, we utilized the $\mathrm{PENCIL\,\,CODE}$ \citep{2001ApJ...550..824B}. The computational domain is a periodic cube with a size of $(2\pi)^3$, discretized into a grid of $400^3$ points. The code solves the magnetohydrodynamic (MHD) equations, given by:
\begin{eqnarray}
\frac{D \rho}{Dt} &=& -\rho \, {\bf \nabla} \cdot {\bf U}, \label{continuity_equation_Code}\\
\frac{D {\bf U}}{Dt} &=& -c_s^2 {\bf \nabla} \ln \rho + \frac{{\bf J} \times {\bf B}}{\rho} + \nu \left({\bf \nabla}^2 {\bf U} + \frac{1}{3} {\bf \nabla} ({\bf \nabla} \cdot {\bf U}) \right) + {\bf f}_\mathrm{kin}, \label{momentum_equation_Code}\\
\frac{\partial {\bf A}}{\partial t} &=& {\bf U} \times {\bf B} + \eta \, {\bf \nabla}^2 {\bf A}, \label{vector_induction_equation_Code}\\
\Rightarrow \frac{\partial {\bf B}}{\partial t} &=& {\bf \nabla} \times ({\bf U} \times {\bf B}) + \eta \, {\bf \nabla}^2 {\bf B}, \label{magnetic_induction_equation_Code}
\end{eqnarray}
where $\rho$ is the density, ${\bf U}$ is the velocity, ${\bf B}$ is the magnetic field, ${\bf A}$ is the vector potential, and ${\bf J}$ represents the current density. The term $D/Dt$ refers to the advective derivative, defined as $\partial / \partial t + {\bf U} \cdot {\bf \nabla}$. The parameter $\eta = c^2 / 4\pi \sigma$ represents the magnetic diffusivity, where $c$ is the speed of light and $\sigma$ is the conductivity.\\

These equations are nondimensionalized. The velocity and magnetic fields are scaled by the sound speed $c_s$ and $(\rho_0 \mu_0)^{1/2} c_s$, respectively, based on the relations $E_\mathrm{M} \sim B^2/\mu_0$ and $E_V \sim \rho_0 U^2$, where $\mu_0$ is the magnetic permeability and $\rho_0$ is the initial density. The plasma is assumed to be weakly compressible, implying $\rho \sim \rho_0$. The forcing term ${\bf f}_\mathrm{kin}(x,t)$ is defined as $N {\bf f}(t) \exp (i {\bf k}_f(t) \cdot {\bf x} + i\phi(t))$, where $N$ is a normalization constant, ${\bf f}$ is the forcing magnitude, and ${\bf k}_f(t)$ is the wave number of the forcing. At each time step, one of 20 random vectors from the ${\bf k}_f$ set is chosen. For simplicity, we set $c_s$, $\mu_0$, and $\rho_0$ to 1, making the equations dimensionless.\\

The forcing function ${\bf f}(t)$ is expressed as $f_0 \mathbf{f}_k(t)$:
\begin{eqnarray}
{\bf f}_k(t) = \frac{i \mathbf{k}(t) \times (\mathbf{k}(t) \times \mathbf{\hat{e}}) - \lambda |{\bf k}(t)| (\mathbf{k}(t) \times \mathbf{\hat{e}})}{k(t)^2 \sqrt{1 + \lambda^2} \sqrt{1 - (\mathbf{k}(t) \cdot \mathbf{\hat{e}})^2/k(t)^2}},
\label{forcing amplitude fk}
\end{eqnarray}
where $k = 2\pi/l$, and $l$ is the scale. Large scales correspond to $k = 1$, while smaller turbulent scales correspond to $k > 2$. The parameter $\lambda = \pm 1$ controls the helicity of the field, generating fully right- or left-handed helical fields. The unit vector $\mathbf{\hat{e}}$ is arbitrary. For our simulation, we used fully helical kinetic energy ($\lambda = 1$) at $k_f \sim 5$. The forcing function appears in Eq.~(\ref{momentum_equation_Code}) with $f_0 = 0.07$ for the helical kinetic forcing dynamo (HKFD). When $\lambda = 0$, the forcing is nonhelical. {It is important to note that the Reynolds averaging, also known as the Reynolds rule \cite{1990cp...book.....M}, is not applied to this energy source, i.e., $\langle f \rangle \neq 0$.}\\

The MHD equations require initial conditions, and a seed magnetic field of $B_0 \sim 10^{-4}$ was introduced. However, this field quickly decays due to the forcing function and the turbulent flow's lack of memory. Initially, small-scale magnetic energy is dominant but decreases as the simulation progresses. The plasma was driven with $\nu = \eta = 0.006$ for $Re_\mathrm{M} = 261$, $\nu = \eta = 0.003$ for $Re_\mathrm{M} = 607$, and $\nu = \eta = 0.0006$ for $Re_\mathrm{M} = 4290$ with the resolution of $400^3$. Here, the magnetic Reynolds number $Re_\mathrm{M}$, which can indicate the level of turbulence in the fluid system, is defined as $-\mathbf{U}\cdot\nabla \mathbf{B} /  \eta \nabla^2\mathbf{B} \sim UL/\eta$, where $U$ is the characteristic velocity and $L\,(\sim \nabla^{-1})$ represents the characteristic length scale of the system ($2\pi$). In all simulations, the magnetic Prandtl number $Pr_\mathrm{M} = \nu/\eta$ was set to 1. By avoiding excess kinetic or magnetic energy near the dissipation scale (which occurs when $Pr_\mathrm{M} \neq 1$), we focused on studying the $\alpha$ and $\beta$ effects on the large-scale magnetic energy $\overline{E}_\mathrm{M}$ and magnetic helicity $\overline{H}_\mathrm{M}$ in the HKFD. The helicity ratio was set to $f_h = 1$ (fully helical), and energy was injected at the forcing scale with $k = 5$.

\begin{figure*}
\centering
\includegraphics[width=16 cm]{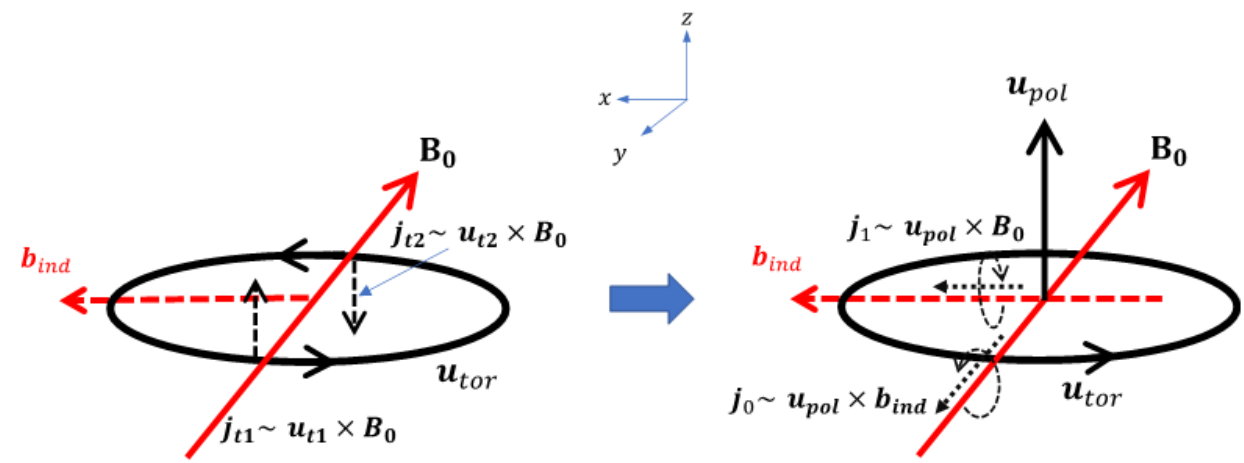}
\caption{
Left: nonhelical kinetic eddy.
Right: positively polarized helical kinetic eddy.
$B_0$ represents the seed magnetic field, and $t$ indicates the transverse direction.
}
\label{f6}
\end{figure*}

\subsection{Theoretical method}

\subsubsection{Conventional Derivation of $\alpha$  and  $\beta$}
With Reynolds rule the large scale magnetic induction equation is represented as
\begin{eqnarray}
\frac{\partial \overline{\bf B}}{\partial t}&=&\nabla \times \langle {\bf u}\times {\bf b}\rangle-\eta \nabla \times \nabla \times \overline{\bf B}.
\label{magnetic_induction_EMF}
\end{eqnarray}
The electromotive force (EMF) $\langle {\bf u}\times {\bf b}\rangle$ is inherently nonlinear, {making exact analytic calculations} challenging. However, the EMF can be approximately linearized using the parameters $\alpha$, $\beta$, and the large-scale magnetic field $\overline{\bf B}$: $\langle \mathbf{u}\times \mathbf{b}\rangle \sim \alpha \overline{\mathbf{B}} - \beta \nabla \times \overline{\mathbf{B}}$. Consequently, the equation can be rewritten as:
\begin{eqnarray}
\frac{\partial \overline{\bf B}}{\partial t}&\sim& \nabla\times (\alpha\overline{\bf B}-\beta \nabla \times \overline{\bf B)}-\eta \nabla \times \nabla \times \overline{\bf B}\nonumber\\
&\sim&\alpha \overline{\bf J} + (\beta+\eta) \nabla^2 \overline{\bf B}.
\label{magnetic_induction_alpha_beta}
\end{eqnarray}
Eq.~(\ref{magnetic_induction_alpha_beta}) suggests that the magnetic field in the plasma system is induced by the current density $\overline{\bf J}$ through the $\alpha$ effect, in accordance with Ampère's law. While this is fundamentally electrodynamic, it can be related to the static Biot-Savart law. Additionally, $\nabla^2 \overline{\bf B}$ indicates the fields are decreased through diffusion, the $\beta$ effect. This arises from the relation $\nabla \times (\nabla \times \overline{\bf B})= - \nabla^2 \overline{\bf B}$.\\

In the Solar dynamo, Eq.~(\ref{magnetic_induction_alpha_beta}) is separated into poloidal and toroidal components as follows \citep{2014ARA&A..52..251C}:
{
\begin{eqnarray}
&&\frac{\partial \overline{A}}{\partial t}=(\eta+\beta)\bigg(\nabla^2-\frac{1}{\zeta^2} \bigg)\overline{A}-\frac{{\bf u}_p}{\zeta}\cdot \nabla (\zeta\overline{A})+\alpha \overline{B}_\mathrm{tor}~(+S(r,~\theta,~\overline{B}_\mathrm{tor})),\label{Solar_poloidal_magnetic_field}\\
&&\frac{\partial \overline{B}_\mathrm{tor}}{\partial t}=(\eta+\beta)\bigg(\nabla^2-\frac{1}{\zeta^2} \bigg)\overline{B}_\mathrm{tor}+\frac{1}{\zeta}\frac{\partial(\zeta\overline{B}_\mathrm{tor})}{\partial r}\frac{\partial (\eta+\beta)}{\partial r}\label{Solar_toroidal_magnetic_field}\\
&&-\zeta{\bf u}_p\cdot \nabla \bigg(\frac{\overline{B}_\mathrm{tor}}{\zeta}\bigg)-\overline{B}_\mathrm{tor}\nabla\cdot{\bf u}_p+\zeta(\nabla \times (\overline{A}\hat{e}_\phi))\cdot\nabla{\Omega}+\nabla\times(\alpha\nabla \times\overline{A}\hat{e}_\phi).
\end{eqnarray}
(Here, $\overline{\mathbf{B}}_{\text{pol}} = \nabla \times \overline{\mathbf{A}}$, $\zeta = r \sin \theta$, and $\mathbf{u}_\mathrm{p} = \mathbf{u}_\mathrm{r} + \mathbf{u}_\mathrm{\theta}$. Additionally, $u_{\text{tor}} = u_\mathrm{\phi} = \Omega r \sin \theta$ is dropped with $u_{\text{tor}} / r$.)}\\
The poloidal magnetic field $\overline{\mathbf{B}}_{\text{pol}} (=\nabla \times \overline{\mathbf{A}})$ is coupled to the toroidal component $\overline{\mathbf{B}}_{\text{tor}}$ through the $\alpha$ effect. The evolution of $\overline{\mathbf{B}}_{\text{pol}}$ is influenced by the $\beta$ effect and the source term $S(r,~\theta,~\overline{B}_{\text{tor}})$ from the Babcock-Leighton model. Meanwhile, $\overline{\mathbf{B}}_{\text{tor}}$ is driven by the $\beta$ effect and differential rotation $\nabla \Omega$. Note that the angular velocity $\Omega$ originates from the convective motion $\overline{\mathbf{U}} = \mathbf{r} \times \Omega$. Jouve et al. \cite{2008A&A...483..949J} simulated these coupled equations. However, the exact definitions of $\alpha$ and $\beta$ remain unknown. Therefore, they tested various models for $\alpha$ and $\beta$ and used the least $\alpha$ that does not diverge the equations. At present, three closure theories derived the fundamental forms of $\alpha$ and $\beta$ with some unknown quantities\citep{1976JFM....77..321P, 2008matu.book.....B}.\\

\subsubsection{Derivation of $\alpha$ and $\beta$}
(i) MFT model \cite{2001ApJ...550..824B, 2008matu.book.....B}\\
Using mean field theory (MFT), the coefficients are given by
\begin{eqnarray}
\alpha_{\mathrm{MFT}} &=&  \frac{1}{3}\int^{\tau} \big(\langle {\bf j}\cdot {\bf b}\rangle - \langle {\bf u}\cdot \nabla\times {\bf u}\rangle\big)\, dt,\label{alpha_beta_MFT}\\
\beta_{\mathrm{MFT}} &=& \frac{1}{3}\int^{\tau} \langle u^2\rangle\, dt, \label{beta_MFT}
\end{eqnarray}
where the exact correlation time $\tau$ is unknown, spatial isotropy is assumed, and higher-order terms are neglected. This approach is the most intuitive and straightforward to use.\\

(ii) DIA model\\
In the Direct Interaction Approximation (DIA, \cite{Akira2011}), the nonlinear higher-order terms are incorporated using the Green function. Additionally, a second-order statistical relation is employed in place of the vector identity typically used in Mean Field Theory (MFT):
\begin{eqnarray}
\langle U_\mathrm{i}({k})U_\mathrm{j}({-k})\rangle = \left( \delta_\mathrm{ij} - \frac{k_\mathrm{i} k_\mathrm{j}}{k^2} \right) E_V(k) + \frac{i}{2} \frac{k_\mathrm{l}}{k^2} \epsilon_\mathrm{ijl} H_V(k), \label{Statistical_relation}
\end{eqnarray}
where $\int E_V(k) d{k} = \langle U^2 \rangle / 2$ and $\int H_V(k) d{k}=\langle {\bf U}\cdot \nabla\times {\bf U} \rangle$. The electromotive force (EMF) is represented by the coefficients $\alpha$, $\beta$, and $\gamma$ for cross helicity $\langle {\bf u}\cdot {\bf b} \rangle$:
\begin{eqnarray}
\alpha_{\mathrm{DIA}} &=& \frac{1}{3} \int d{\bf k} \int^t G \left( \langle {\bf j}\cdot {\bf b} \rangle - \langle {\bf u}\cdot \nabla\times {\bf u} \rangle \right) d\tau, \label{DIA_alpha} \\
\beta_{\mathrm{DIA}} &=& \frac{1}{3} \int d{\bf k} \int^t G \left( \langle u^2 \rangle + \langle b^2 \rangle \right) d\tau, \label{DIA_beta} \\
\gamma_{\mathrm{DIA}} &=& \frac{1}{3} \int d{\bf k} \int^t G \langle {\bf u}\cdot {\bf b} \rangle \, d\tau, \label{DIA_gamma}
\end{eqnarray}

(iii) EDQNM model\\
With additional differentiation over time, momentum equation and magnetic induction equation yield the fourth order terms. In EDQNM, the fourth-order moments $\langle x_\mathrm{l} x_m x_n x_q \rangle$ are replaced by second-order moments, i.e., $\sum_\mathrm{lmnq}\langle x_\mathrm{l}x_m \rangle\langle x_nx_q \rangle$. This process, quasi-normalization, is known to be valid in the isotropic and homogeneous system. The second order moments are represented with $E_V(E_\mathrm{M})$, $H_V(H_\mathrm{M})$, and cross helicity $\langle{\bf u}\cdot {\bf b}\rangle$ (see Eq.~(\ref{Statistical_relation}), \citep{1967PhFl...10..859K, 1976JFM....77..321P, 1990cp...book.....M}).  $\alpha$ and $\beta$ are represented as:
\begin{eqnarray}
\alpha_{\mathrm{QN}}&=&\frac{2}{3}\int^{t} \Theta_{\mathrm{kpq}}(t)\big(\langle {\bf j}\cdot {\bf b} \rangle- \langle {\bf u}\cdot \nabla\times {\bf u}\rangle\big)\,dq,
\label{EDQNM_alpha}\\
\beta_{\mathrm{QN}}&=&\frac{2}{3}\int^{t} \Theta_{\mathrm{kpq}}(t)\langle u^2 \rangle\,dq.
\label{EDQNM_beta}
\end{eqnarray}
Here, the relaxation time $\Theta_\mathrm{kpq}$ is defined as $\frac{1 - \exp(-\mu_\mathrm{kpq}t)}{\mu_\mathrm{kpq}}$, which converges to a constant value over time: $\Theta_\mathrm{kpq} \rightarrow \mu^{-1}_\mathrm{kpq}$. The eddy damping operator $\mu_\mathrm{kpq}$ is determined experimentally. Note that the coefficients of $\alpha$ and $\beta$ are $2/3$, which arises from the quasi-normalization of fourth-order moments into the combinations of second-order moments.\\

Compared to MFT and EDQNM, DIA includes the $\gamma$ effect, which is associated with cross helicity. Technically, this effect can also be produced using the EDQNM approach or also possibly with MFT. However, {the derivation of cross helicity is required} longer and more complex calculations. At present, Kraichnan \cite{1967PhFl...10..859K} for the nonhelical dynamo and Pouquet et al. \cite{1976JFM....77..321P} for the helical dynamo did not include the related term. Additionally, $\beta$ in DIA includes the effect of turbulent magnetic energy $b^2$ as well as turbulent kinetic energy $u^2$.\\

\subsubsection{Alternative Approach to $\alpha$  and  $\beta$ with $\overline{E}_\mathrm{M}$  and  $\overline{H}_\mathrm{M}$}
In addition to the conventional method based on closure theories, we introduce an indirect approach to determine the profiles of $\alpha$ and $\beta$ using the fundamental quantities $\overline{E}_\mathrm{M}$ and $\overline{H}_\mathrm{M}$, rather than relying on ambiguous turbulent data. We have previously employed and developed this model \citep{2020ApJ...898..112P,  2023ApJ...944....2P, 2025PhRvD.111b3021P}. However, for its application to real data or DNS, it is important to explain the key concepts and derivations instead of merely referencing prior works. Furthermore, this analytic approach is continuously being refined and applied to various physical cases.\\

From Eq.~(\ref{magnetic_induction_alpha_beta}), we derive the coupled equations for $\overline{H}_\mathrm{M}(t)$ and $\overline{E}_\mathrm{M}(t)$ as follows:
\begin{eqnarray}
\frac{\partial}{\partial t} \big( \overline{\mathbf{A}} \cdot \overline{\mathbf{B}} \big)
&=& 2\alpha \overline{\mathbf{B}} \cdot \nabla \times \overline{\mathbf{A}} + 2(\beta + \eta) \overline{\mathbf{B}} \cdot \nabla^2 \overline{\mathbf{A}} \nonumber \\
\Rightarrow \frac{\partial \overline{H}_\mathrm{M}}{\partial t} &=&  4\alpha \overline{E}_\mathrm{M} - 2(\beta + \eta)\overline{H}_\mathrm{M}, \label{Hm1} \\
\frac{\partial}{\partial t} \big( \overline{\mathbf{B}} \cdot \overline{\mathbf{B}} \big)
&=& 2\alpha \overline{\mathbf{J}} \cdot \overline{\mathbf{B}} + 2(\beta + \eta) \overline{\mathbf{B}} \cdot \nabla^2 \overline{\mathbf{B}} \nonumber \\
\Rightarrow \frac{\partial \overline{E}_\mathrm{M}}{\partial t} &=&  \alpha \overline{H}_\mathrm{M} - 2(\beta + \eta)\overline{E}_\mathrm{M}. \label{Em1}
\end{eqnarray}
Here, we have used the substitution $\nabla \rightarrow i k$ with $k = 1$ for the large-scale field. Therefore, instead of the current helicity $\overline{H}_C = \langle \mathbf{J} \cdot \mathbf{B} \rangle = k^2 \langle \mathbf{A} \cdot \mathbf{B} \rangle = k^2 \overline{H}_\mathrm{M}$, we can express it in terms of the magnetic helicity. This represents a special case of a large-scale helical dynamo. In general, $H_C \neq H_\mathrm{M}$. These coupled differential equations can then be solved in the usual manner:
\begin{eqnarray}
\left[
\begin{array}{c}
\frac{\partial \overline{H}_\mathrm{M}}{\partial t} \\
\frac{\partial \overline{E}_\mathrm{M}}{\partial t}
\end{array}
\right]
&=&\left[
\begin{array}{cc}
-2(\beta+\eta) & 4\alpha\\
     \alpha         &-2(\beta+\eta)
\end{array}
\right]
\left[
\begin{array}{c}
\overline{H}_\mathrm{M} \\
\overline{E}_\mathrm{M}
\end{array}
\right]
=
\left[
\begin{array}{cc}
\lambda & 0\\
0 &\lambda
\end{array}
\right]
\left[
\begin{array}{c}
\overline{H}_\mathrm{M} \\
\overline{E}_\mathrm{M}
\end{array}
\right],\\
&\Rightarrow& \lambda_\mathrm{1,\, 2}=\pm2\alpha-2(\beta+\eta),\quad X = \frac{1}{\sqrt{5}}\left[\begin{array}{cc} 2&2\\ 1&-1 \end{array} \right]
\label{MatrixofHMandEm}
\end{eqnarray}
The solutions are
\begin{eqnarray}
\left[
\begin{array}{cc}
{\overline H}_\mathrm{M}(t) \\
{\overline E}_\mathrm{M}(t)
\end{array}
\right]=\frac{1}{\sqrt{5}}
\left[
\begin{array}{cc}
2c_1e^{\int^t \lambda_1 d\tau}+2c_2e^{\int^t \lambda_2 d\tau} \\
c_1e^{\int^t \lambda_1 d\tau}-c_2e^{\int^t \lambda_2 d\tau}
\end{array}
\right].
\label{SolutionofHMandEm}
\end{eqnarray}
With the initial conditions for $c_1=H_\mathrm{M0}$ and $c_2=E_\mathrm{M0}$, we have
\begin{eqnarray}
2\overline{H}_\mathrm{M}(t)&=&(2\overline{E}_\mathrm{M0}+\overline{H}_\mathrm{M0})e^{2\int^{t}_0(\alpha-\beta-\eta)d\tau}-(2\overline{E}_\mathrm{M0}-\overline{H}_\mathrm{M0})e^{2\int^{t}_0(-\alpha-\beta-\eta)d\tau},\label{HmSolutionwithAlphaBeta1}\\
4\overline{E}_\mathrm{M}(t)&=&(2\overline{E}_\mathrm{M0}+\overline{H}_\mathrm{M0})e^{2\int^{t}_0(\alpha-\beta-\eta)d\tau}+(2\overline{E}_\mathrm{M0}-\overline{H}_\mathrm{M0})e^{2\int^{t}_0(-\alpha-\beta-\eta)d\tau}.\label{EmSolutionwithAlphaBeta2}
\end{eqnarray}
These solutions satisfy the realizability condition $2\overline{E}_\mathrm{M}>\overline{H}_\mathrm{M}$. Additionally, for right handed helical kinetic forcing (or left handed helical magnetic forcing), $\alpha\sim \langle \mathbf{j}\cdot \mathbf{b} \rangle - \langle \mathbf{u}\cdot \omega \rangle $ becomes negative. Then, the second terms in {each solution} are dominant, leading to $2\overline{H}_\mathrm{M}(t)=-(2\overline{E}_\mathrm{M0}-\overline{H}_\mathrm{M0})e^{2\int^{t}_0(-\alpha-\beta-\eta)d\tau}$, $4\overline{E}_\mathrm{M}(t)=(2\overline{E}_\mathrm{M0}-\overline{H}_\mathrm{M0})e^{2\int^{t}_0(-\alpha-\beta-\eta)d\tau}$. The magnetic helicity ratio on large scale is $\overline{f}_\mathrm{h}=2\overline{H}_\mathrm{M}(t)/4\overline{E}_\mathrm{M}(t)\rightarrow +1$. This result was confirmed by the values of $f_\mathrm{hm}$ at $k=1$ in Figs.~\ref{f2a_new}, \ref{f2b_new}, \ref{f3b}, \ref{f3d}, \ref{f3f}. If the system is forced with left handed helical kinetic energy or right handed helical magnetic energy, the opposite result is obtained.\\

To find \(\alpha\) and \(\beta\), we can either use direct substitution or apply a simple trick. Here, we introduce the simpler method. By multiplying Eq.~(\ref{Em1}) by 2 and subtracting it from Eq.~(\ref{Hm1}), we obtain an equation for \(\overline{H}_\mathrm{M} - 2\overline{E}_\mathrm{M}\). Conversely, by adding the two equations, we derive another equation for \(\overline{H}_\mathrm{M} + 2\overline{E}_\mathrm{M}\). From these, we can proceed to derive the desired expressions:
\begin{eqnarray}
\alpha_{\mathrm{semi}}(t)&=&\frac{1}{4}\frac{d}{dt}log_e \bigg|\frac{ 2\overline{E}_\mathrm{M}(t)+\overline{H}_\mathrm{M}(t)}{2\overline{E}_\mathrm{M}(t)-\overline{H}_\mathrm{M}(t)}\bigg|,\label{alphaSolution3}\\
\beta_{\mathrm{semi}}(t)&=&-\frac{1}{4}\frac{d}{dt}log_e\big| \big(2\overline{E}_\mathrm{M}(t)-\overline{H}_\mathrm{M}(t) \big)\big( 2\overline{E}_\mathrm{M}(t)+\overline{H}_\mathrm{M}(t)\big)\big|-\eta,\label{betaSolution3}
\label{betaSolution31}
\end{eqnarray}
We also need to verify the results with numerically simulated data before applying them to real-world data. To obtain the profiles, a data set of \(\overline{E}_\mathrm{M}(t)\) and \(\overline{H}_\mathrm{M}(t)\) from direct numerical simulations (DNS) with time intervals is required. We used an approximation such as
\[
\frac{\Delta \overline{E}_\mathrm{M}}{\Delta t} \sim \frac{\overline{E}_\mathrm{M}(t_\mathrm{n}) - \overline{E}_\mathrm{M}(t_\mathrm{n-1})}{t_\mathrm{n} - t_\mathrm{n-1}}.
\]
The various \(\alpha\) and \(\beta\) profiles shown in Figs.~4, 5, and 6 were generated using this method with DNS data. This approach provides the reference profiles for determining \(\alpha\) and \(\beta\). Moreover, it requires only large-scale magnetic data, which are relatively easy to measure. However, the method has a significant limitation when the large-scale helical magnetic field becomes saturated, i.e., \(2\overline{E}_\mathrm{M} \sim \pm \overline{H}_\mathrm{M}\) (see Figs.~\ref{f2a}, \ref{f2c}, \ref{f2e}), \ref{f2c_new}, \ref{f2d_new}. In such cases, the logarithmic function is not well defined as the argument approaches zero. $\alpha$ and $\beta$ depend on the difference between $|\overline{H}_\mathrm{M}|$ and $\overline{E}_\mathrm{M}$. {Below is a portion of our IDL code script for $\alpha$ and  $\beta$ in Eq.~(\ref{alphaSolution3}), (\ref{betaSolution3}) from the large scale magnetic energy ($spec\_mag$(i, j)) and helicity ($spechel\_mag$(i, j))}:
\begin{verbatim}
for j=0L,  i_last-1 do begin
    c[j]=2.0*spec_mag(1, j) + spechel_mag(1, j)   % k=1
    d[j]=2.0*spec_mag(1, j) - spechel_mag(1, j)
endfor

for j=0L,  i_last-1 do begin
  alpha_semi[j] = 0.25*[(ALOG(c[j+1])-ALOG(c[j]))
             -(ALOG(d[j+1])-ALOG(d[j]))]/(time[j+1]-time[j])
   beta_semi[j] =-0.25*((ALOG(c[j+1])-ALOG(c[j]))
             +(ALOG(d[j+1])-ALOG(d[j])))/(time[j+1]-time[j])-eta
endfor
\end{verbatim}
Here, `spec\_mag', `spechel\_mag' are from pencil\_code power spectrum data for magnetic energy and magnetic helicity in Fourier space.

\subsection{Derivation of $\beta$ with $E_\mathrm{V}$ and $H_\mathrm{V}$}
{In the previous section, we introduced a method to determine the $\alpha$ and $\beta$ coefficients using large-scale magnetic field data. This approach is applicable regardless of whether the data come from DNS, observations, or experiments, as it relies solely on the largest-scale magnetic structures. As a result, it avoids ambiguities in measurement and accurately reflects the $\alpha$ and $\beta$ profiles generated by various effects within the system. However, this method does not explain the underlying mechanisms in the small-scale, turbulent regions of the system that give rise to such profiles. Therefore, by deriving $\alpha$ and $\beta$ using theoretical approaches that consider various physical effects in the turbulent region and comparing the resulting profiles, we can gain clearer insights into the internal generation mechanisms. In this paper, we first focus on the $\beta$ effect. Traditional approaches have closed the equations by approximating the second-order velocity moment with only the kinetic energy. However, when kinetic helicity is present, including its contribution leads to more accurate results.}\\

Now, we check the possibility of negative $\beta$ using analytic method.
\begin{eqnarray}
&&\langle {\bf u}\times \int^{\tau}(-{\bf u}\cdot \nabla \overline{\bf B})dt \rangle \rightarrow \big\langle -\epsilon_\mathrm{ijk} u_\mathrm{j} (r)u_m(r+l)\tau \frac{\partial \overline{B}_k}{\partial \overline{r}_m}\big\rangle \label{beta_derivation_helical_first}\\
&\sim&-\epsilon_\mathrm{ijk}\langle u_\mathrm{j}u_m\rangle\frac{\partial \overline{B}_k}{\partial \overline{r}_m}-\langle u_\mathrm{j}\,l_n\partial_n u_m\rangle\epsilon_\mathrm{ijk}\frac{\partial \overline{B}_k}{\partial \overline{r}_m}\\
&\sim&\underbrace{-\frac{1}{3}\langle u^2\rangle\epsilon_\mathrm{ijk}\frac{\partial \overline{B}_k}{\partial \overline{r}_m}\delta_\mathrm{jm}}_1\,\underbrace{-\big \langle \frac{l}{6}|H_V|\big \rangle\epsilon_\mathrm{ijk}\frac{\partial
\overline{B}_k}{\partial \overline{r}_m}\delta_\mathrm{nk}\delta_\mathrm{mi}}_2,\label{beta_derivation_helical1}
\end{eqnarray}
Here, we set \(\tau \rightarrow 1\) for simplicity, assuming that the two eddies are correlated over one eddy turnover time. We apply a more general identity for the second-order moment as follows \citep{1990cp...book.....M, 2008tufl.book.....L}:
\begin{eqnarray}
U_\mathrm{jm}\equiv\langle u_\mathrm{j}(r)u_m(r+l)\rangle=A(l)\delta_\mathrm{jm}+B(l)l_\mathrm{j}l_m+C(l)\epsilon_\mathrm{jms}l_s.
\label{general_beta_derivation1}
\end{eqnarray}
With the reference frame of $\vec{l}=(l,\,0,\,0)$ or any other appropriate coordinates, we can easily infer the relation of $A$, $B$, and $C$ as follows: $A+l^2B\equiv F$, $A\equiv G$, $(U_\mathrm{23}=)lC\equiv H$. Then, Eq.(\ref{general_beta_derivation1}) is represented as
\begin{eqnarray}
U_\mathrm{jm}=G\,\delta_\mathrm{jm}+\frac{(F-G)}{l^2}\,l_\mathrm{j}l_m+H\epsilon_\mathrm{jms}\frac{l_\mathrm{s}}{l}.
\label{general_beta_derivation2}
\end{eqnarray}
With the incompressibility condition $\nabla \cdot {\bf U}=0$, we get the additional constraint.
\begin{eqnarray}
\frac{\partial U_\mathrm{jm}}{\partial l_\mathrm{j}}=\frac{l_\mathrm{j}}{l}G'\delta_\mathrm{jm}+4l_m\frac{F-G}{l^2}+l_m\frac{(F'-G')l^2-2l(F-G)}{l^3}=0,
\label{constraint_F_G}
\end{eqnarray}
which leads to $G=F+(l/2)\,\partial F/\partial l$. So, the second order moment is
\begin{eqnarray}
U_\mathrm{jm}=\bigg(F+\frac{l}{2}\frac{\partial F}{\partial l}\bigg)\delta_\mathrm{jm}-\frac{l}{2l^2}\frac{\partial F}{\partial l}l_\mathrm{j}l_m+H\epsilon_\mathrm{jms}\frac{l_s}{l}.
\label{general_beta_derivation3}
\end{eqnarray}
If $j=m$, $U_\mathrm{jj}=F=u^2/3 =E_V/6$. And, if $j\neq m$, the relation $\langle \epsilon_\mathrm{ijk} u_\mathrm{j} (r)u_m(r+l)\partial \overline{B}_k/\partial \overline{r}_m\rangle\rightarrow -\langle \epsilon_\mathrm{ijk}l_\mathrm{j}l_m/2l\, \partial F /\partial l \rangle\partial \overline{B}_k/\partial \overline{r}_m$ implies that any $m$ makes the average negligible. And, for $H$, we use Lesieur's approach \citep{2008tufl.book.....L}:
\begin{eqnarray}
H_V&=&\lim_\mathrm{y\rightarrow x} \bf u(x)\cdot \nabla\times {\bf u(y)}\nonumber\\
&=&\lim_\mathrm{y\rightarrow x}\epsilon_\mathrm{ijn} u_\mathrm{i}\frac{\partial u_n}{\partial y_\mathrm{j}} =\lim_\mathrm{l\rightarrow 0}\epsilon_\mathrm{ijn} \frac{\partial U_\mathrm{in}(l)}{\partial l_\mathrm{j}}\,\,\,(\leftarrow y=x+l)\nonumber\\
&=&\lim_\mathrm{l\rightarrow 0}\epsilon_\mathrm{ijn} \epsilon_\mathrm{ins}\bigg(\delta_\mathrm{js}\frac{H}{l}-\frac{l_\mathrm{j}l_\mathrm{s}}{l^3}H+\frac{l_\mathrm{j}l_\mathrm{s}}{l^2}\frac{\partial H}{\partial l}\bigg)=-\frac{6}{l}H
\label{general_beta_derivation4}
\end{eqnarray}


Then, $U_\mathrm{jm}$ is
\begin{eqnarray}
U_\mathrm{jm}=\frac{\langle u^2\rangle}{3}\delta_\mathrm{jm}-\epsilon_\mathrm{jms}\frac{l_\mathrm{s}}{6}H_V.
\label{general_beta_derivation5}
\end{eqnarray}
EMF by the advection term $-{\bf u}\cdot \nabla \overline{\bf B}$ is
\begin{eqnarray}
\big\langle -\epsilon_\mathrm{ijk} u_\mathrm{j} (r)u_\mathrm{m}(r+l)\frac{\partial \overline{B}_\mathrm{k}}{\partial \overline{r}_\mathrm{m}}\big\rangle&=&
-\frac{1}{3}\langle u^2\rangle \epsilon_\mathrm{ijk}\frac{\partial \overline{B}_\mathrm{k}}{\partial \overline{r}_\mathrm{m}}\delta_\mathrm{jm}+
\epsilon_\mathrm{ijk}\epsilon_\mathrm{jms}\frac{l_\mathrm{s}}{6}H_\mathrm{V}\frac{\partial \overline{B}_\mathrm{k}}{\partial \overline{r}_\mathrm{m}}\nonumber\\
&\Rightarrow&-\frac{1}{3}\langle u^2\rangle \nabla\times \overline{\bf B}+\frac{l}{6}H_\mathrm{V} \big(\nabla \times \overline{\bf B}\big)
\label{general_beta_derivation6}
\end{eqnarray}
For the second term in RHS, we referred to vector identity $\epsilon_\mathrm{ijk}\epsilon_\mathrm{jms}=\delta_\mathrm{km}\delta_\mathrm{is}-\delta_\mathrm{ks}\delta_\mathrm{im}\rightarrow -\delta_\mathrm{ks}\delta_\mathrm{im}$ with the consideration of $\nabla\cdot \overline{\bf B}=0$.
\begin{eqnarray}
\big\langle \epsilon_\mathrm{ijk}\epsilon_\mathrm{jms}\frac{l_s}{6}H_\mathrm{V}\frac{\partial \overline{B}_\mathrm{k}}{\partial \overline{r}_m}\big\rangle \rightarrow \langle \epsilon_\mathrm{jik}\frac{l}{6}H_\mathrm{V}\rangle \,\epsilon_\mathrm{ijk} \frac{\partial \overline{B}_\mathrm{k}}{\partial \overline{r}_\mathrm{i}}\rightarrow \frac{l}{6}H_\mathrm{V} \big(\nabla \times \overline{\bf B}\big)_\mathrm{j}.
\label{general_beta_derivation7}
\end{eqnarray}
We used the normal permutation rule and regarded $l_s$ as $l$. Finally,
\begin{eqnarray}
\big( \frac{1}{3}\langle u^2\rangle - \frac{l}{6}H_\mathrm{V} \big)\big(-\nabla \times \overline{\bf B}\big)\equiv \beta \big(-\nabla \times \overline{\bf B}\big).
\label{general_beta_derivation8}
\end{eqnarray}
We infer the constraint of `$l$':
\begin{eqnarray}
\langle u(r)u(r+l)\rangle\equiv g(r)\langle u^2(r)\rangle\sim \frac{1}{3}\langle u^2\rangle-\frac{l}{6}\langle {\bf u}\cdot \nabla\times {\bf u}\rangle=\frac{2}{3}E_\mathrm{V}-\frac{l}{6}H_\mathrm{V}.
\label{new_beta_derivation_helical3}
\end{eqnarray}
For negative $g(r)$, which is a typical property of parallel correlation function \citep{2004tisebook.....D}, is $l>4E_\mathrm{V}/H_\mathrm{V}$.

Kinetic helicity that constitutes $\beta$ here is a pseudo scalar, which undergoes a phase reversal under reflection. However, when combined with the correlation length component $l_\mathrm{s}$, they exhibit the characteristics of a general scalar. When the parity operator is applied to each\citep{2005mmp..book.....A}, the signs of $l_\mathrm{s}$ and $H_\mathrm{V}$ change, but the overall sign of $l_\mathrm{s} \cdot H_\mathrm{V}$ remains unchanged. Since there is no change in $\langle u(r)u(r\pm l)\rangle$ either, the general scalar nature is preserved overall. Such characteristics are not uncommon in turbulence models, and both Eq.~(\ref{Statistical_relation}) in Fourier space and Eq. (\ref{general_beta_derivation1}), (\ref{general_beta_derivation2}) in real space share the same form.

\subsection{Numerical verification of $\alpha$ and $\beta$}
In Fig.~7, these coefficients were used to reproduce the evolving EMF and large-scale magnetic field. \(\alpha_\mathrm{\text{semi}}\) was obtained with large-scale magnetic data \(\overline{E}_\mathrm{M}(t)\) and \(\overline{H}_\mathrm{M}(t)\) according to Eq.~(\ref{alphaSolution3}). For \(\beta\), two kinds of magnetic diffusivity, \(\beta_\mathrm{\text{semi}}\) and \(\beta_\mathrm{\text{theo}}\), were obtained and compared. \(\beta_\mathrm{\text{semi}}\) was derived from the large-scale magnetic data according to Eq.~(\ref{betaSolution3}); however, \(\beta_\mathrm{\text{theo}}\) was obtained with the turbulent kinetic data \(E_V(t)\) and \(H_V(t)\) in small scale [Eq.~(\ref{general_beta_derivation8})]. The left panel in Fig.~7 illustrates that the EMF with \(\alpha\) and \(\beta\) is generally consistent with the EMF from DNS. However, they diverge as the fields become saturated. Typically, the EMF with \(\beta_\mathrm{\text{theo}}\) is larger than that from DNS, indicating that \(\beta_\mathrm{\text{theo}}\) is not quenched sufficiently. The differences among \(\beta_\mathrm{\text{DNS}}\), \(\beta_\mathrm{\text{semi}}\), and \(\beta_\mathrm{\text{theo}}\) are observed in the large-scale magnetic fields as they become saturated in the right panel. With these \(\alpha\) and \(\beta\), we reproduced the large-scale magnetic field using a simple numerical method:  $\Delta \overline{B}(t) / \Delta t \sim \nabla \times \alpha \overline{B}(t) + (\beta + \eta)\nabla^2 \overline{B}(t)$. The IDL script for \(\overline{B}(t)\) is
\begin{verbatim}
B[0] =  sqrt(2.0*spec_mag(1, 0))  % k=1

for j=0L,  t_last do begin
  B[j+1] = B[j] + (-alpha[j]-beta[j]-eta)*B[j]*(time[j+1]-time[j])
  % helical magnetic field
endfor
\end{verbatim}
Here, we have used $-\alpha[j]$ based on $\nabla \times \alpha \overline{B} = -\alpha \overline{B}$ for the right handed helical kinetic forcing, as illustrated in Eq.~(\ref{forcing amplitude fk}) ($\lambda=+1$), Figs.~\ref{f2a_new}, \ref{f2b_new}. The combination of $\beta_\mathrm{\mathrm{semi}}$ with $\alpha_\mathrm{\mathrm{semi}}$ reproduces the DNS result relatively well. However, $\overline{B}$ from $\beta_\mathrm{\mathrm{theo}}$ shows some discrepancies near the saturation regime, consistent with the left panel. $\beta_\mathrm{\mathrm{theo}}$, which is based on the hydrodynamic tensor identity [Eq.~(\ref{general_beta_derivation2})], is not sufficiently quenched in the regime where the magnetic field strengthens. In MHD, where the velocity (magnetic) field is an implicit function of the magnetic (velocity) field, these implicit terms may need to be represented explicitly. In DIA, magnetic energy is included in $\beta$ as shown in Eq.~(\ref{DIA_beta}). However, DIA separates the counter-kinematic approach (magnetic approach) from the kinematic approach and independently uses the magnetic second moment $\langle b(r)b(r+l)\rangle$ \cite{Akira2011}. However, it is unclear whether the implicit effects can be incorporated into the tensor identity at present. A more thorough study may be required.

\section{Result}
Figs.~\ref{f1a}-\ref{f1f} show the temporal evolution of root mean squared velocity $U_\mathrm{rms}$ and magnetic field $B_\mathrm{rms}$ (left panel), and the spectra of kinetic energy $E_V$ and magnetic energy $E_\mathrm{M}$ for magnetic Reynolds numbers $Re_\mathrm{M} \sim 261$, 607, and 4290 (right panel). Both $U_\mathrm{rms}$ and $B_\mathrm{rms}$ were calculated as $U_\mathrm{rms} = \sqrt{2 \int E_V \, dk}$ and $B_\mathrm{rms} = \sqrt{2 \int E_\mathrm{M} \, dk}$, respectively. The corresponding magnetic diffusivities and velocities for different $Re_\mathrm{M}$ are $\eta = 6 \times 10^{-3}$ with $U_\mathrm{rms} = 0.25$, $\eta = 3 \times 10^{-3}$ with $U_\mathrm{rms} = 0.29$, and $\eta = 6 \times 10^{-4}$ with $U_\mathrm{rms} = 0.41$. Since the two larger $Re_\mathrm{M}$ cases have not yet reached saturation, $U$ was approximately estimated for these cases. The evolution and saturation profiles differ slightly across the cases. The left panel shows that as $Re_\mathrm{M}$ increases, more time is required for the fields to reach saturation. Additionally, the right panel indicates that as $Re_\mathrm{M}$ increases, the small-scale regime of the energy spectrum flattens. Despite these variations, some common features are observed for the unit magnetic Prandtl number, $Pr_\mathrm{M}$. First, the left panel consistently shows that $U$ decreases as $B$ grows, demonstrating the conversion of kinetic energy into magnetic energy. Second, increasing $Re_\mathrm{M}$ leads to stronger saturated fields. Furthermore, the increasing $Re_\mathrm{M}$ flattens the energy spectrum, transitioning from $k^{-12/3}$ to $k^{-6/3}$ near to Kolmogorov's scaling factor, $k^{-5/3}$. This behavior reflects the influence of increased small-scale energy on enhancing diffusion effects.\\

Figs.~\ref{f2a}-\ref{f2f} present the spectra of magnetic helicity $H_\mathrm{M} = \langle \mathbf{A} \cdot \mathbf{B} \rangle$ and magnetic energy $2E_\mathrm{M} = \langle B^2 \rangle$ in the left panel, and kinetic helicity $H_V = \langle \mathbf{v} \cdot \omega \rangle$ alongside kinetic energy $2E_V = \langle U^2 \rangle$ in the right panel. The values of $\nu$ and $\eta$ are consistent with those in Fig.~1. As shown in the left panel, both $\overline{H}_\mathrm{M}$ and $\overline{E}_\mathrm{M}$ reveal that magnetic strength at $k=1$ increases over time, eventually exceeding that at the forcing scale $k=5$. In contrast, magnetic strength at smaller scales ($k > 1$) diminishes as $\overline{B}$ at $k=1$ grows, clearly demonstrating the inverse cascade of magnetic energy. {Note that the polarity of $H_\mathrm{M}$ at large scales is negative in this right-handed kinetic forcing system.}. Additionally, Figs. \ref{f1e} and \ref{f1f} exhibit Kolmogorov's energy scaling. The different scaling factors—$H_\mathrm{M} \sim k^{-1}B^2 \sim k^{-3.1}$, $H_V \sim kU^2 \sim k^{-1.3}$, $E_\mathrm{M} \sim k^{-6/3}$, and $E_V \sim k^{-6/3}$—remain consistent, despite their small discrepancies. These variations stem from the limited Reynolds number and the relatively small resolution $400^3$. On the right panel, the evolution of plasma motion shows a dependence on the Reynolds number. For instance, comparing Fig. \ref{f1e} with Fig. \ref{f1f}, $E_V$ and $H_V$ converge by $t \sim 50$, faster than the magnetic field's evolution. Although the same viscosity and diffusivity were used ($\nu = \eta = 0.006$, $Pr_\mathrm{M} = 1$), their asymmetric evolutions might seem unusual. However, Eqs.~(\ref{momentum_equation_Code}) and (\ref{magnetic_induction_equation_Code}) are inherently not symmetric, and the coupling between them adds further complexity to their behavior. As the Reynolds number increases, the asymmetrical balance between $E_V$ and $E_\mathrm{M}$ becomes more pronounced.\\

Figs.~\ref{f2a_new} and \ref{f2b_new} show the spectra of current helicity, $\langle \mathbf{J} \cdot \mathbf{B} \rangle$ and $k\langle B^2 \rangle$, with the aim of demonstrating the conservation of magnetic helicity. Due to the small magnitude of magnetic helicity at small scales, we instead use current helicity, $k^2 H_m$. The system is driven with positive (right-handed) helical kinetic energy at $k = 5$. This forcing generates magnetic helicity of opposite polarization, i.e., left-handed magnetic helicity, which is inversely cascaded toward large scale. To conserve total magnetic helicity, right-handed magnetic helicity is produced at smaller scales. This explains why the polarities of the large-scale and small-scale regimes are opposite. Figs.~\ref{f2c_new} and \ref{f2d_new} show the temporal evolution of large-scale magnetic energy, $2\overline{E}_\mathrm{M}$, and large-scale magnetic helicity, $|\overline{H}_\mathrm{M}|$. As noted earlier, magnetic helicity in large scale is negative, but we use its absolute value for comparison. Initially, $|\overline{H}_\mathrm{M}|$ is smaller than $2\overline{E}_\mathrm{M}$ in both cases. However, their magnitudes converge over time, such that $|\overline{H}_\mathrm{M}| \rightarrow 2\overline{E}_\mathrm{M}$. The discrepancy $2\overline{E}_\mathrm{M}-|\overline{H}_\mathrm{M}|$ is related to the determination of $\alpha$ and $\beta$.\\

Figs.~\ref{f3a}-\ref{f3f} include the large-scale kinetic energy (\(10\times \overline{E}_V\), black dot-dashed line) and magnetic energy (\(10\times \overline{E}_\mathrm{M}\), red solid line) in the left panel. Since $\overline{E}_V$ is very small throughout the entire range, we have multiplied both $\overline{E}_V$ and $\overline{E}_\mathrm{M}$ by 10 for better visibility. In the right panel, the kinetic helicity ratio \(f_\mathrm{hk} = \langle \mathbf{U} \cdot (\nabla \times \mathbf{U}) \rangle / (k \langle U^2 \rangle)\) and the magnetic helicity ratio \(f_\mathrm{hm} = k \langle \mathbf{A} \cdot \mathbf{B} \rangle / \langle B^2 \rangle\) are illustrated. \(\overline{E}_\mathrm{M}\) increases at \(t \sim 200{-}400\) and becomes saturated. In contrast, \(\overline{E}_V\) remains at a very low value. We also include the evolving profiles of the \(\alpha\) (dotted line) and \(\beta\) (dashed line) coefficients, obtained from the large-scale magnetic energy and magnetic helicity. The \(\alpha\) coefficient oscillates between positive and negative values, while \(\beta\) remains negative. Both coefficients converge to zero as \(\overline{E}_\mathrm{M}\) saturates. However, the contribution of \(\alpha\) is much smaller than conventional inference, which predicts it should remain negative and converge to zero as the magnetic field grows (see \(\alpha \sim \int \big(\langle \mathbf{j}\cdot \mathbf{b} \rangle - \langle \mathbf{u} \cdot \omega \rangle \big) d\tau\), where \(\langle \mathbf{u} \cdot \omega \rangle > \langle \mathbf{j} \cdot \mathbf{b} \rangle\rightarrow \langle \mathbf{u} \cdot \omega \rangle \sim \langle \mathbf{j} \cdot \mathbf{b} \rangle\)). With the Laplacian \(\nabla^2 \rightarrow -k^2\), the negative \(\beta\) amplifies the magnetic field: $\partial \overline{B} / \partial t = \alpha \overline{J} + (\beta +\eta) \nabla^2 \overline{B}$. The \(\beta\) effect originates from the spatial distribution of the curl of the current density, $-\nabla \times \overline{J} = \nabla^2 \overline{B}$.  Mathematically, the \(\beta\) effect here represents diffusion in plasmas. Plasma turbulence, acting as a diffusion mechanism, can efficiently transport magnetic energy to larger scales. Near saturation, \(\beta\) oscillates heavily during this nonlinear stage. We applied a smoothing function in IDL, averaging over 10 nearby points. The $\alpha$ and $\beta$ profiles were obtained from the evolving magnetic data $\overline{B}$. Regardless of the theory behind $\alpha$ and $\beta$, it must account for these profiles. We will soon demonstrate that the $\alpha$ and $\beta$ coefficients, based on these profiles, can reproduce the evolving $\overline{B}$ obtained from DNS.\\

The evolution of the helicity ratios is consistent with a system forced by helical right-handed kinetic energy. The kinetic helicity ratio \(f_\mathrm{hk}\) at \(k=5\) (the forcing scale) remains close to 1 (with \(\lambda = +1\), see Eq.~[\ref{forcing amplitude fk}]). Other \(f_\mathrm{hk}\) values at \(k=1\) for the large scale and \(k=8\) for the small-scale regime show positive oscillations. In contrast, the magnetic helicity ratios \(f_\mathrm{hm}\) evolve quite differently. The large-scale helicity ratio eventually converges to \(-1\), while those at smaller scales grow to positive values, indicating the conservation of magnetic helicity. We will discuss this issue further using vector analysis and the field structure model. \(\Delta t\) indicates the simulation time interval between each data point.\\

Figs.~\ref{f4a}-\ref{f4f} compare the profiles of \(\alpha\) and \(\beta\). The \(\alpha\) profile (represented as \(\alpha_\mathrm{\mathrm{EM-HM}}\) or \(\alpha_\mathrm{\mathrm{semi}}\), dot-dashed line) in the left panel is calculated using the large-scale magnetic data \(\overline{E}_\mathrm{M}\) and \(\overline{H}_\mathrm{M}\). The other lines are obtained from $ \alpha_\mathrm{\mathrm{old}}\equiv \tau/3 \left( \langle \mathbf{j} \cdot \mathbf{b} \rangle - \langle \mathbf{u} \cdot \boldsymbol{\omega} \rangle \right)$, where the correlation time \(\tau\) is set to one unit of simulation time. Here, \(\langle \mathbf{j} \cdot \mathbf{b} \rangle\) and \(\langle \mathbf{u} \cdot \boldsymbol{\omega} \rangle\) represent the current helicity and kinetic helicity in the small scales, respectively. This \(\alpha_\mathrm{\text{old}}\) is a basic form of MFT, DIA, and EDQNM. However, the exact range of the turbulent or small-scale regime for these pseudo-scalars is not yet known. Therefore, we compared them for different ranges: \(k = 2{-}4\) (dotted line), \(k = 2{-}6\) (solid), and \(k = 2{-}k_\mathrm{\text{max}}\) (dashed). The value of \(\alpha_\mathrm{\text{old}}\) for \(k = 2{-}4\) is negligibly small, while the other ranges including the forcing scale $k=5$ yield practically the same results. Apparently, kinetic helicity at the forcing scale largely looks to determine the profile of \(\alpha_\mathrm{\text{old}}\). However, as shown in Fig.~\ref{f4e}, the \(\alpha_\mathrm{\text{old}}\) profiles for \(k = 2{-}6\) and \(k = 2{-}k_\mathrm{\text{max}}\) are distinct. This separation suggests that more helicity migrates toward the smaller scale regime with increasing $Re_\mathrm{M}$. On the other hand, the \(\alpha\) profile derived from the large-scale magnetic data oscillates and converges to zero much earlier. The relatively weak \(\alpha\) effect indicates a small electromagnetic contribution to the plasma dynamo. Statistically, in a plasma system composed of many charged particles constraining one another, the strong electromagnetic effect is somewhat limited. The weak \(\alpha\) effect is quite reasonable in such plasmas. Nevertheless, it is this electrodynamic effect that determines the system's polarity in the early time regime.\\

The right panel includes the profile of \(\beta\) (\(\beta_\mathrm{\text{EM-HM}}\) or \(\beta_\mathrm{\text{semi}}\), dot-dashed line) derived from large-scale magnetic data, alongside \(\beta_\mathrm{\text{theo}}\) calculated from turbulent kinetic energy \(\langle u^2 \rangle\) and kinetic helicity \(\langle \mathbf{u} \cdot \nabla \times \mathbf{u} \rangle\). While conventional theory attributes kinetic helicity solely to the \(\alpha\) effect, detailed calculations reveal that it also influences \(\beta\) diffusion. For a non-helical velocity field, \(\beta\) remains positive, consistent with traditional theory. However, in the presence of helicity, the helical kinetic structure can diffuse magnetic energy to larger scales. In Fig.~\ref{f4b}, we compare the conventional \(\beta_\mathrm{\text{old}}\) with \(\tau/3 \langle u^2 \rangle\) (\(k = 2 - k_\mathrm{\text{max}}\), \(\tau \rightarrow 1\), dotted line), \(\beta_\mathrm{\text{EM-HM}}\), and \(\beta_\mathrm{\text{theo}}\) across various ranges. The plot shows that \(\beta_\mathrm{\text{old}}\) remains positive, diffusing the large-scale magnetic field. In contrast, \(\beta_\mathrm{\text{theo}}\) and \(\beta\) (indicating \(\beta_\mathrm{\text{EM-HM}}\)) become negative, potentially amplifying the large-scale magnetic field. The plot demonstrates that \(\beta_\mathrm{\text{theo}}\) generally aligns with \(\beta_\mathrm{\text{EM-HM}}\), except in the early and saturated time regimes (logarithmic time scale). In the early regime, the effect of the initial seed magnetic field persists. In the saturated regime, turbulent magnetic effects seem to play a more significant role in the \(\beta\) profile. The range of consistency is relatively extensive. Since \(\alpha_\mathrm{\text{EM-HM}}\) and \(\beta_\mathrm{\text{EM-HM}}\) serve as standard references for \(\overline{B}(t)\), the alignment of \(\beta_\mathrm{\text{theo}}\) indicates that turbulent kinetic energy and kinetic helicity contribute to \(\beta\) diffusion. However, near saturation, the discrepancies between \(\beta_\mathrm{\text{EM-HM}}\) and \(\beta_\mathrm{\text{theo}}\) in Figs.~\ref{f4b}, \ref{f4d}, and \ref{f4f} suggest that additional magnetic effects are needed to fully account for the quenching of the \(\beta\) effect.\\

Figs.~\ref{f4a_new}-\ref{f4c_new} illustrate the ratio of \(\beta_\mathrm{\text{theo}}\) to \(\beta_\mathrm{\text{semi}}\). As the plots show, while \(Re_\mathrm{M}\) is small, these two coefficients are quite close. As \(Re_\mathrm{M}\) increases, the contribution of \(\beta_\mathrm{\text{theo}}\) surpasses the actual \(\beta_\mathrm{\text{semi}}\). This indicates that \(\beta_\mathrm{\text{theo}}\), composed of the turbulent velocity field, is not quenched sufficiently, implying that \(\beta\) diffusion requires additional effects. One of the good candidates is the turbulent magnetic effect.\\

Figs.~\ref{f5a}-\ref{f5f} demonstrate the turbulent electromotive force (EMF, \(\langle \mathbf{u} \times \mathbf{b} \rangle\), left panel) and the evolution of the large-scale magnetic field \(\overline{B}(t)\) in the right panel. These plots verify the validity of the newly obtained \(\alpha\) and \(\beta\) coefficients from Fig.~4. In the left panel, we compare the curl of \(\langle \mathbf{u} \times \mathbf{b} \rangle\) with its linearized representations, \(\alpha \overline{B} - \beta \nabla \times \overline{B}\). As mentioned earlier, the exact range of the turbulent scale remains unclear. Therefore, instead of directly calculating \(\nabla \times \langle \mathbf{u} \times \mathbf{b} \rangle\), we replaced it with \(\partial \overline{\mathbf{B}} / \partial t - \eta \nabla^2 \overline{\mathbf{B}}\) (red solid line). For \(\alpha\) and \(\beta\), however, direct calculation is possible. We included two types of \(\alpha \overline{B} - \beta \nabla \times \overline{B}\) for comparison. The semi-\(\alpha\) and semi-\(\beta\) (dotted line) represent \(\alpha\) and \(\beta\) obtained from large-scale magnetic data, while the semi-\(\alpha\) and theo-\(\beta\) (dot-dashed line) indicate \(\alpha\) from the large-scale magnetic data and \(\beta_\mathrm{\text{theo}}\) derived from the small-scale kinetic data.\\

The plots show that the EMF with \(\alpha_\mathrm{\mathrm{semi}}\) and \(\beta_\mathrm{\mathrm{semi}}\), as well as the EMF with \(\alpha_\mathrm{\mathrm{semi}}\) and \(\beta_\mathrm{\mathrm{theo}}\), coincide with the DNS EMF \(\nabla \times \langle \mathbf{u} \times \mathbf{b} \rangle\) in the kinematic regime. However, these \(\alpha\) and \(\beta\) EMFs begin to diverge slightly before the DNS EMF becomes saturated. The EMF with semi-\(\alpha\) and semi-\(\beta\) evolves lower than the DNS EMF, while the EMF with semi-\(\alpha\) and theo-\(\beta\) evolves higher than the DNS EMF. All \(\alpha\) or \(\beta\) from large-scale magnetic data oscillate significantly in the nonlinear regime. These coefficients require the calculation of \(2\overline{E}_\mathrm{M} - \overline{H}_\mathrm{M}\) for the logarithmic argument, which becomes very small near the saturated regime (\(2\overline{E}_\mathrm{M} \sim \overline{H}_\mathrm{M}\) in Eqs.~(\ref{alphaSolution3}), (\ref{betaSolution3}), left panel in Fig.~2). The difference between them exceeds the computational roundoff error, leading to significant numerical noise. To mitigate this, a smoothing function in IDL (averaging over 10 neighboring points) was applied, but it proved insufficient.\\

In the right panel, the evolution of the large-scale magnetic field \(\overline{\mathbf{B}}\) is shown. \(\overline{\mathbf{B}}\) from DNS (black solid line), \(\overline{\mathbf{B}}\) from semi-\(\alpha\) and semi-\(\beta\) (dot-dashed line), and \(\overline{\mathbf{B}}\) from \(\alpha_\mathrm{\mathrm{semi}}\) and \(\beta_\mathrm{\mathrm{theo}}\) (red dashed line) are presented. The reproduced fields coincide quite well with the DNS data. However, as the dynamo process becomes nonlinear, the theoretical \(\beta\) as well as $\beta_\mathrm{\mathrm{semi}}$ becomes less efficient. These evolutions are consistent with the field profiles in Fig.~5. \(\beta_\mathrm{\mathrm{theo}}\) from turbulent kinetic data closely matches \(\beta_\mathrm{\mathrm{semi}}\) from large-scale magnetic data, except in the early time regime and the saturated regime. In the early time regime, the effect of the initial conditions persists, while in the nonlinear regime, \(\beta_\mathrm{\mathrm{theo}}\) does not fully capture the growth of the turbulent magnetic field. This discrepancy in the nonlinear regime leads to differences in the profiles of \(\overline{\mathbf{B}}\). \(\overline{\mathbf{B}}\) from the conventional $\alpha$ and $\beta$ is much smaller and out of range.\\

\section{Discussion}
As $Re_\mathrm{M}$ increases, the magnetic field becomes frozen into the fluid eddies, allowing its intrinsic properties to emerge. Fig.~5 shows that the $\alpha$ effect is not dominant, while the $\beta$ effect becomes more significant. Considering the evolution of large-scale magnetic fields (Fig.~4, left panel), $\beta$ quenching appears to be a more critical factor than $\alpha$ quenching in the growth of helical magnetic fields. Compared to the conventional theory of $\alpha$ quenching, this is a major difference we have discovered in this work.\\

To understand the physical meanings of $\alpha$ and $\beta$ for EMF, we consider the field structure shown in Fig.~8. The left figure represents a counterclockwise circular structure of a plasma turbulence eddy, over which a magnetic field \(\mathbf{B}_0\) threads through the eddy, influencing its structure and energy distribution. The plasma motions, \(\mathbf{u}_\mathrm{t1}\) and \(\mathbf{u}_\mathrm{t2}\), with \(\mathbf{B}_0\) to produce current densities, \(j_\mathrm{t1}\hat{z}\) and \(-j_\mathrm{t2}\hat{z}\). According to Ampère's law, these two current densities induce a magnetic field, \(\mathbf{b}_\mathrm{ind}\hat{x}\). In other words, the curl of \(\mathbf{j}_\mathrm{t1}\) and \(\mathbf{j}_\mathrm{t2}\) can expressed as \(\nabla \times \mathbf{j}_\mathrm{t} = \nabla \times (\nabla \times \mathbf{b}_\mathrm{ind}) = -\nabla^2 \mathbf{b}_\mathrm{ind}\), representing the diffusion of the magnetic field. Eventually, $B_0$ and $u_t$  should weaken in the presence of this induced magnetic field.\\

If the velocity field is helical, however, we have an additional velocity poloidal component \(\mathbf{u}_\mathrm{\text{pol}}\), as shown in the right figure. This velocity interacts with the induced magnetic field to generate current density: \(\mathbf{u}_\mathrm{\text{pol}} \times \mathbf{b}_\mathrm{\text{ind}} \rightarrow \mathbf{j}_\mathrm{0}\). The current density \(j_\mathrm{0}\hat{y}\) induces an additional magnetic field \(b_\mathrm{\text{tor}}\), which forms left-handed magnetic helicity with \(\mathbf{B}_\mathrm{0}\). Here, \(\mathbf{B}_\mathrm{0}\) acts as the poloidal field, and \(b_\mathrm{\text{tor}}\) acts as the toroidal field. The toroidal and poloidal fields strengthen each other in what is called the \(\alpha^2\) process. Additionally, \(\mathbf{u}_\mathrm{\text{pol}}\) can interact with \(\mathbf{B}_\mathrm{0}\) to generate \(\mathbf{j}_\mathrm{1}\). In this case, \(\mathbf{b}_\mathrm{\text{ind}}\) and \(\mathbf{j}_\mathrm{1}\) form right-handed magnetic helicity on a smaller scale. However, the interaction \({\bf u}_\mathrm{pol}\times \mathbf{b}_\mathrm{\text{ind}}\) appears to dominate over \({\bf u}_\mathrm{pol}\times \mathbf{B}_\mathrm{0}\) because $u_\mathrm{pol}$ and $B_0$ do not generally align. This net process repeats, producing magnetic helicities that add or subtract pseudo-scalars. Figs. 3, ~\ref{f3b}, \ref{f3d}, and \ref{f3f} ($f_\mathrm{hk}$, $f_\mathrm{hm}$ at $k=1,~5,~8$) are associated with this process and are based on the linearization of the EMF, \(\langle \mathbf{u} \times \mathbf{b} \rangle \rightarrow \alpha \mathbf{B} - \beta \nabla \times \mathbf{B}\) and $\beta\sim E_V-lH_V$ (Eq.~[\ref{general_beta_derivation8}]).\\

On the other hand, $E_V$ is a normal scalar, whereas $H_V$ is a pseudoscalar. Under the parity operator (reflection symmetry \citep{2005mmp..book.....A, 2009LNP...780.....S}), $H_V$ reverses its sign, while $E_V$ remains unchanged. However, the quantity $l$ also changes its sign under the parity operator. {Consequently, $lH_V$ behaves as an ordinary scalar.} The second order velocity moment in Eq.~(\ref{Statistical_relation}) is an example of a statistical identity involving both a normal scalar and a pseudoscalar. Eq.~(\ref{general_beta_derivation5}) represents a second-order statistical velocity identity in real space \citep{2008tufl.book.....L}. In addition, there exist statistical identities corresponding to the magnetic field, which have the same mathematical form and share the same physical properties. These identities play a crucial role in closing the equations in EDQNM \citep{1976JFM....77..321P}.

\section{Conclusions}
In this paper, we calculated $\alpha$ and $\beta$ by increasing $\text{Re}_\mathrm{M}$ while keeping $\text{Pr}_\mathrm{M}$ fixed at 1. We then reproduced and compared $\overline{B}$ using $\alpha$ and $\beta$. The values of $\alpha_\mathrm{\text{semi}}$ and $\beta_\mathrm{\text{semi}}$ calculated from large-scale magnetic $\overline{H}_\mathrm{M}$ and $\overline{E}_\mathrm{M}$ accurately reproduced $\overline{B}$, except for the case where $\overline{H}_\mathrm{M} \sim 2\overline{E}_\mathrm{M}$. This indicates that $\alpha_\mathrm{\text{semi}}$ and $\beta_\mathrm{\text{semi}}$ can serve as a reference for observations, simulations, and theoretical analyses, provided that the system has not yet reached full saturation. However, since they are obtained from $\overline{H}_\mathrm{M}$ and $\overline{E}_\mathrm{M}$, which contain all physical effects, it is necessary to derive them analytically using the original theoretical method, namely the function iterative method with a closure theory, to understand the underlying physical mechanisms.\\

Here, we replaced the second moment of the turbulence velocity, $\langle u(r)u(r+l) \rangle$, with turbulent kinetic energy and kinetic helicity to derive and apply $\beta_\mathrm{\text{theo}}$. Unlike conventional theories, the effect of kinetic helicity can make the second moment value negative, demonstrating that magnetic diffusion is directed toward $\overline{B}$. In other words, when the poloidal fluid motion $u_\mathrm{\text{pol}}$ combines with the toroidal fluid motion $u_\mathrm{\text{tor}}$ and the magnetic field, the direction of diffusion can change. We briefly examined this mechanism in Fig.~8. Negative magnetic diffusion is not a forbidden specific phenomenon; rather, it becomes positive when combined with the Laplacian $\nabla^2$, leading to energy migration toward larger scales. Hydrodynamics also suggests effects that imply negative diffusion. Specifically, for two eddies aligned in the same direction, $\langle u_y u_y (r+l) \rangle \sim u^2 g(r)$, a region where $g(r) < 0$ can appear momentarily as $r \rightarrow \infty$\citep{2004tisebook.....D}. However, the impact is not significant, suggesting that the interaction between the magnetic field and $u_\mathrm{\text{pol}}$ plays a more prominent role.\\

Meanwhile, {we determined the correlation length $l$} through a trial-and-error method, but the relationship between this length and {the strength of kinetic helicity in the system} remains unclear. Furthermore, as the effect of the magnetic field increases, $\beta_\mathrm{\text{theo}}$ fails to accurately represent $\overline{B}$. This implies that $\langle u(r)u(r+l) \rangle$ should include magnetic field effects in addition to turbulent kinetic energy and kinetic helicity. However, since the direct relationship between $u$ and $b$ is currently unclear, it is difficult to explain how to modify the identity of the second moment. This appears to require further research.\\

Overall, the small $\alpha$ effect and the large $\beta$ effect in magnetic field amplification explain how the magnetic field is induced in plasma. In systems where charged particles interfere with other magnetic eddies, it shows that the effect of magnetic diffusion due to fluid characteristics is greater than electromagnetic effects. However, this does not imply that the $\alpha$ effect is insignificant. As seen in Fig.~8, the $\alpha$ effect appears to determine the system's polarity in conjunction with the conservation of magnetic helicity, while the actual amplification of the magnetic field is likely governed by $\beta$ diffusion. As $Re$ or $Re_\mathrm{M}$, or the degree of nonlinearity, increases, the fluidic characteristics of plasma, due to its electrical neutrality in plasmas, become more dominant. Furthermore, in certain regions, the enhanced magnetic field naturally evolves in a way that lowers the energy density by increasing the eddy scale, following the law of entropy increase. Finally, in this paper, we examined the case where $\text{Pr}_\mathrm{M} = 1$; however, to accurately describe natural phenomena, it is also necessary to investigate the general ranges where $\text{Pr}_\mathrm{M} \gg 1$ and $\text{Pr}_\mathrm{M} \ll 1$. The study on $\alpha$ and $\beta$ presented in this work aims to reproduce the DNS and observed magnetic fields to obtain accurate profiles and, from these, understand the physical mechanisms of the internal processes that give rise to $\alpha$ and $\beta$.

\section*{Acknowledgements}
The author acknowledges the support from the physics department at Soongsil University.

\bibliographystyle{apsrev4-2}
\bibliography{bibfile0202} 








\label{lastpage}
\end{document}